\documentclass[pra,reprint,letterpaper,nofootinbib,showpacs,floatfix]{revtex4-1}
\usepackage{graphicx}
\usepackage{dcolumn}
\usepackage{amsmath}
\usepackage{hyperref}
\usepackage{algorithm}
\usepackage[noend]{algpseudocode}

\hypersetup{
    bookmarksnumbered=true, % If Acrobat bookmarks are requested, include section numbers
    unicode=false, % non-Latin characters in Acrobat bookmarks
    pdfstartview={FitH}, % fits the width of the page to the window
    pdftitle={}, % title
    pdfauthor={}, % author
    pdfsubject={}, % subject of the document
    pdfcreator={}, % creator of the document
    pdfproducer={}, % producer of the document
    pdfkeywords={}, % list of keywords
    pdfnewwindow=true, % links in new window
    colorlinks=true, % false: boxed links; true: colored links
    linkcolor=blue, % color of internal links
    citecolor=blue, % color of links to bibliography
    filecolor=blue, % color of file links
    urlcolor=blue % color of external links
}
\newcommand{\eq}[1]{(\hyperref[eq:#1]{\ref*{eq:#1}})}

\renewcommand{\sec}[1]{\hyperref[sec:#1]{Section~\ref*{sec:#1}}}
\newcommand{\thrm}[1]{\hyperref[thm:#1]{Theorem~\ref*{thm:#1}}}
\newcommand{\lemm}[1]{\hyperref[lemm:#1]{Lemma~\ref*{lemm:#1}}}
\newcommand{\prop}[1]{\hyperref[prop:#1]{Proposition~\ref*{prop:#1}}}
\newcommand{\corr}[1]{\hyperref[corr:#1]{Corollary~\ref*{corr:#1}}}
\newcommand{\fig}[1]{\hyperref[fig:#1]{Figure~\ref*{fig:#1}}}

\DeclareMathOperator{\tr}{Tr}

\DeclareMathOperator{\CNOT}{CNOT}
\DeclareMathOperator{\sign}{sgn}

\usepackage[table,x11names]{xcolor}

\begin{document}

\title{Validating quantum computers using randomized model circuits}
\author{Andrew W. Cross}
\email{awcross@us.ibm.com}
\author{Lev S. Bishop}
\email{lsbishop@us.ibm.com}
\author{Sarah Sheldon}
\author{Paul D. Nation}
\author{Jay M. Gambetta}
\affiliation{IBM T. J. Watson Research Center, Yorktown Heights, NY 10598}

\begin{abstract}
We introduce a single-number metric, quantum volume, that can be measured using a concrete protocol on near-term quantum computers of modest size ($n\lesssim 50$), and measure it on several state-of-the-art transmon devices, finding values as high as 16. The quantum volume is linked to system error rates, and is empirically reduced by uncontrolled interactions within the system. It quantifies the largest random circuit of equal width and depth that the computer successfully implements. Quantum computing systems with high-fidelity operations, high connectivity, large calibrated gate sets, and circuit rewriting toolchains are expected to have higher quantum volumes. The quantum volume is a pragmatic way to measure and compare progress toward improved system-wide gate error rates for near-term quantum computation and error-correction experiments.
\end{abstract}

\maketitle

Recent quantum computing efforts have moved beyond controlling a few qubits, and are now focused on controlling systems with several tens of qubits \cite{QX, Friis18,Song17}. In these noisy intermediate-scale quantum (NISQ) systems \cite{preskill-nisq}, performance of isolated gates may not predict the behavior of the system. Methods such as randomized benchmarking~\cite{magesan_characterizing_2012}, state and process tomography~\cite{paris_quantum_2004}, and gateset tomography~\cite{merkel_self-consistent_2013,*blume-kohout_demonstration_2017} are valued for measuring the performance of operations on a few qubits, yet they fail to account for errors arising from interactions with spectator qubits \cite{McKay17, Takita17}. Given a system such as this, whose individual gate operations have been independently calibrated and verified, how do we measure the degree to which the system performs as a general purpose quantum computer? We address this question by introducing a single-number metric, the {\em quantum volume}, together with a concrete protocol for measuring it on near-term systems.  Similar to how LINPACK~\cite{LINPACK} and improved benchmarks~\cite{HPCG,HPGMG}, are used for comparing diverse classical computers, this metric is not tailored to any particular system, requiring only the ability to implement a universal set of quantum gates. With the concept of this metric being discussed elsewhere \cite{qvolume,Moll18}, our focus here is on measuring this metric in near-term quantum devices.

The quantum volume protocol we present is strongly linked to gate error rates, and is influenced by underlying qubit connectivity and gate parallelism. It can thus be improved by moving toward the limit in which large numbers of well-controlled, highly coherent, connected, and generically programmable qubits are manipulated within a state-of-the-art circuit rewriting toolchain. High-fidelity state preparation and readout are also necessary. In this work, we evaluate the quantum volume of current IBM Q devices \cite{QX}, and corroborate the results with simulations of the same circuits under a depolarizing error model. While we focus on transmon devices, the protocol can be implemented with any universal programmable quantum computing device.

The quantum volume is based on the performance of random circuits with a fixed but generic form. It is well-known that quantum algorithms can be expressed as polynomial-sized quantum circuits built from two-qubit unitary gates~\cite{NC}. Quantum algorithms are generally not random circuits. However, random circuits model generic state preparations, and are used as the basis of proposals for demonstrating quantum advantage \cite{B18}. In addition, circuits with a similar form appear in near-term algorithms like quantum adiabatic optimization algorithms \cite{FGG14} and variational quantum eigensolvers~\cite{McClean16,*Yung14}.

\begin{figure}[t]
	\includegraphics[width=3.2in]{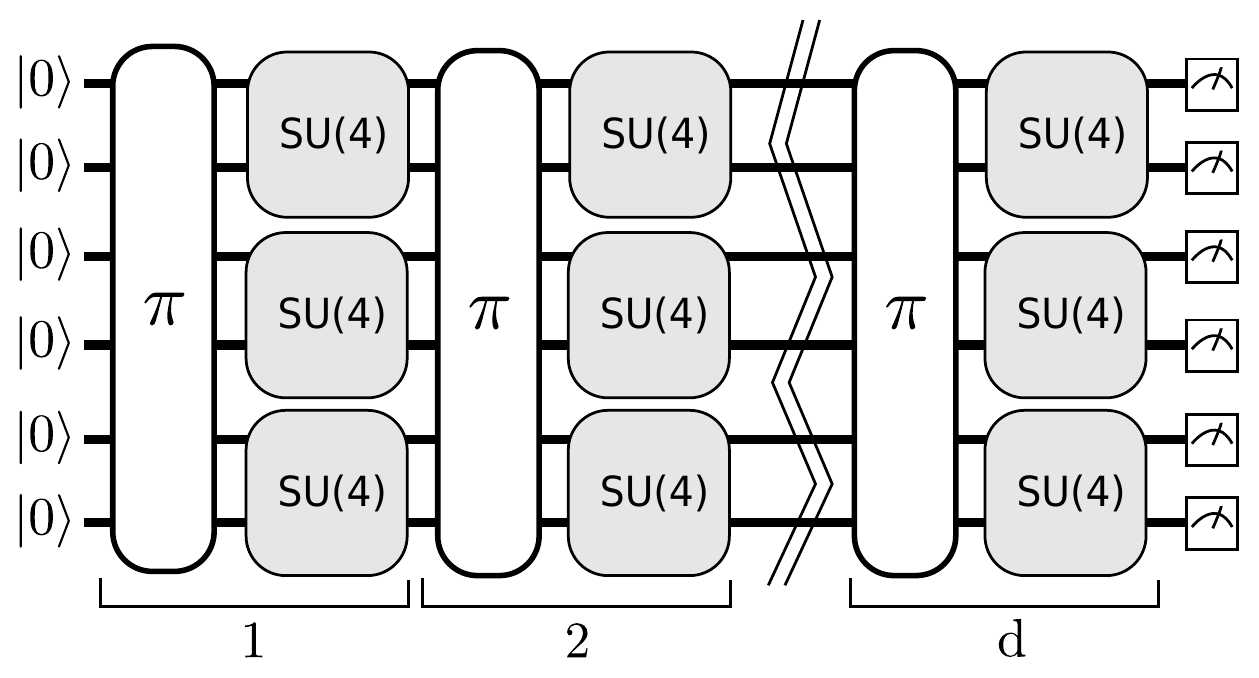}
	\caption{\textbf{Model circuit.} A model circuit consists of $d$ layers of random permutations of the qubit labels, followed by random two-qubit gates. When the circuit width $m$ is odd, one of the qubits is idle in each layer. A final permutation can be applied to the labels of the measurement outcomes. \label{fig:mc}}
\end{figure}

A {\em model circuit}, shown in Fig.~\ref{fig:mc}, with depth $d$ and width $m$, is a sequence $U=U^{(d)}\dots U^{(2)}U^{(1)}$ of $d$ layers
\begin{equation}
U^{(t)} = U^{(t)}_{\pi_t(m'-1),\pi_t(m')}\otimes\dots\otimes U^{(t)}_{\pi_t(1), \pi_t(2)},
\end{equation}
each labeled by times $t=1, \dots, d$ and acting on $m'=2\lfloor n/2\rfloor$ qubits. Each layer is specified by choosing a uniformly random permutation $\pi_t\in S_m$ of the $m$ qubit indices and sampling each $U^{(t)}_{a,b}$, acting on qubits $a$ and $b$, from the Haar measure on SU(4).

To define when a model circuit $U$ has been successfully implemented in practice, we use the {\em heavy output} generation problem \cite{AC16}. The ideal output distribution is
\begin{equation}
p_U(x) = |\langle x|U|0\rangle|^2
\end{equation}
where $x\in \{0,1\}^m$ is an observable bit-string. Consider the set of output probabilities given by the range of $p_U(x)$ sorted in ascending order $p_0\leq p_1\dots\leq p_{2^m-1}$. The median of the set of probabilities is $p_{med} = (p_{2^{(m-1)}} + p_{2^{(m-1)}-1})/2$, and the heavy outputs are
\begin{equation}
H_U = \bigl\{x\in \{0,1\}^m\ \textrm{such that}\ p_U(x) > p_{med}\bigr\}.
\end{equation}
The heavy output generation problem is to produce a set of output strings such that more than two-thirds are heavy. The expected heavy output probability for an ideal device is asymptotically $(1+\ln 2)/2 \sim 0.85$ \cite{AC16}, while it falls to $\sim 0.5$ if the device is completely depolarized.

To evaluate heavy output generation, we implement model circuits using the gate set provided by the target system. For example, the model circuit may need to be rewritten, not only to use the system's gate set, but also to respect the set of available interactions, which may require additional operations such as SWAP gates. The average gate fidelity \cite{horodecki_general_1999} between $m$-qubit unitaries $U$ and $U'$ is
\begin{equation}
F_\text{avg}(U, U') = \frac{\bigl|\mathrm{Tr}(U^\dag U')\bigr|^2/2^m+1}{2^m+1}.
\end{equation}
Given a model circuit $U$, a circuit-to-circuit transpiler finds an implementation $U'$ for the target system such that $1-F_\text{avg}(U,U')\le\epsilon\ll1$. In many cases, the approximation error $\epsilon$ is limited by the selected classical precision within the transpiler (eg. for arithmetic to compute new gate angle parameters), but may be further increased if the hardware requires SU(4) to be approximated with a discrete set of available gates.

The transpiler is free to use all available tricks and hardware resources to implement $U'$ (e.g., taking great computational effort in finding an optimized $U'$, using extra qubits for gate teleportation or temporary storage, etc.). It may optimize over qubit placements by choosing the best region of the device. If it is practical to calibrate a very large gate set, and it happens to include an accurate implementation of $U$, the transpiler is free to use it. None of these approaches is expected to provide an asymptotic advantage, but may significantly improve practical performance. We do require that the transpiler make an honest attempt to implement $U$, and not merely choose a relatively simple operation far from $U$ that nevertheless produces the heavy outputs for $U$. The compilation routine for computing the quantum volume of IBM Q devices is described in Appendix~\ref{app:transpiler}, and an approximation scheme given in Appendix~\ref{app:approximate}. 

The observed distribution for an implementation $U'$ of model circuit $U$ is $q_U(x)$, and the probability of sampling a heavy output is
\begin{equation}
h_U = \sum_{x\in H_U} q_U(x).
\end{equation}
To determine if a given output is heavy, we compute $H_U$ directly from $U$ using a method that scales exponentially\footnote{For error rates as low at $10^{-4}$, we anticipate that model circuits $U$ that can be successfully implemented will involve few enough qubits and/or low enough depth to compute $H_U$ classically. For lower error rates than this, the quantum volume can be superseded by new volume metrics or modified so classical simulations are not necessary.} with $m$. The probability of observing a heavy output by implementing a randomly selected depth $d$ model circuit is $h_d = \int_{U} h_U dU$.  Ideally, we would estimate this quantity using all of the qubits of a large device, but NISQ devices have appreciable error rates, so we begin with small model circuits and progress to larger ones. We are interested in the achievable model circuit depth $d(m)$ for a given model circuit width $m\in [n]$. We define the achievable depth $d(m)$ to be the largest $d$ such that we are confident $h_d>2/3$ (See Appendix~\ref{app:confidence} for further discussion of confidence intervals). In other words,
\begin{equation}
h_1, h_2, \dots, h_{d(m)} > 2/3\ \mathrm{and}\ h_{d(m)+1} \leq 2/3 .
\end{equation}
Algorithm~\ref{alg:qvol} provides pseudocode for testing when each $h_d > 2/3$.

\begin{algorithm}[H]
	\caption{Check heavy output generation}\label{alg:qvol}
	\begin{algorithmic}
		\Function{isHeavy}{$m,d; n_c\ge100,n_s$}
		\State $n_h\gets0$
		\For{$n_c$ repetitions}
		\State $U\gets$ random model circuit, width $m$, depth $d$
		\State $H_U\gets$ heavy set of $U$ from classical simulation
		\State $U'\gets$ compiled $U$ for available hardware 
		\For{$n_s$ repetitions}
		\State $x\gets$ outcome of executing $U'$ 
		\If{$x\in H_U$} 
		$n_h\gets n_h+1$
		\EndIf
		\EndFor
		\EndFor
		\State \Return $\frac{n_h-2\sqrt{n_h(n_s-n_h/n_c)}}{n_c n_s}>\frac{2}{3}$
		\EndFunction
	\end{algorithmic}
\end{algorithm}

We desire a metric that is a single real number, as this enables straightforward comparison. Data $\{d(m)\}$ can be gathered by sweeping over values of $m$ and $d$. We are free to choose any function of this data $\{d(m)\}$ to capture how well a device performs.  The quantum volume treats the width and depth of a model circuit with equal importance and measures the largest square-shaped (i.e., $m=d$) model circuit a quantum computer can implement successfully on average \cite{qvolume,Moll18}. We define the quantum volume $V_Q$ as
\begin{equation}
\log_2 V_Q = \underset{m}{\operatorname{argmax}}  \min(m, d(m))
\end{equation}
and take this definition going forward.

This definition differs from \cite{qvolume,Moll18} and loosely coincides with the complexity of classically simulating the model circuits. There are different ways to classically simulate the model quantum circuits. A straightforward wave-vector propagation approach requires exponential space and time $\sim2^m$. A `Feynman' algorithm uses linear space $\sim d m$ but exponential time $\sim4^{d m}$. It is possible to trade off time and space complexity in a smooth way~\cite{AC16}. Clever partitioning of circuits can achieve good parallelism and efficient use of distributed memory resources for particular supercomputer architectures~\cite{chen_64-qubit_2018,chen_classical_2018,li_quantum_2018,pednault_breaking_2017,boixo_simulation_2017,haner_high_2016,smelyanskiy_qhipster:_2016}. Particular efforts for circuit partitioning and parallelism have been expended for circuits defined on a 2-dimensional square grid of qubits, where the state-of-the-art is $d=40$ for a $9\times9$ grid~\cite{chen_classical_2018}.

One view of these methods is that they use heuristics to approach optimal variable elimination ordering for a tensor network calculation on the graph corresponding to the circuit. The time complexity scales exponentially with the treewidth of the circuit graph~\cite{markov_simulating_2008}. The treewidth is upper-bounded by $m$, and while there are specific circuits of depth $d=4$ with expander graph structure for which the treewidth is $\Omega(m)$, heuristic estimation of the treewidth for some classes of random circuits~\cite{pednault_breaking_2017,boixo_simulation_2017} indicates that the treewidth grows roughly as $d$. Therefore, we heuristically bound the treewidth of the model circuits as $\min(d,m)$, and since the simulation complexity grows exponentially with the treewidth, we define the quantum volume as $V_Q=2^{\min(d,m)}$.

\begin{figure}[h]
	\includegraphics[width=3.4in]{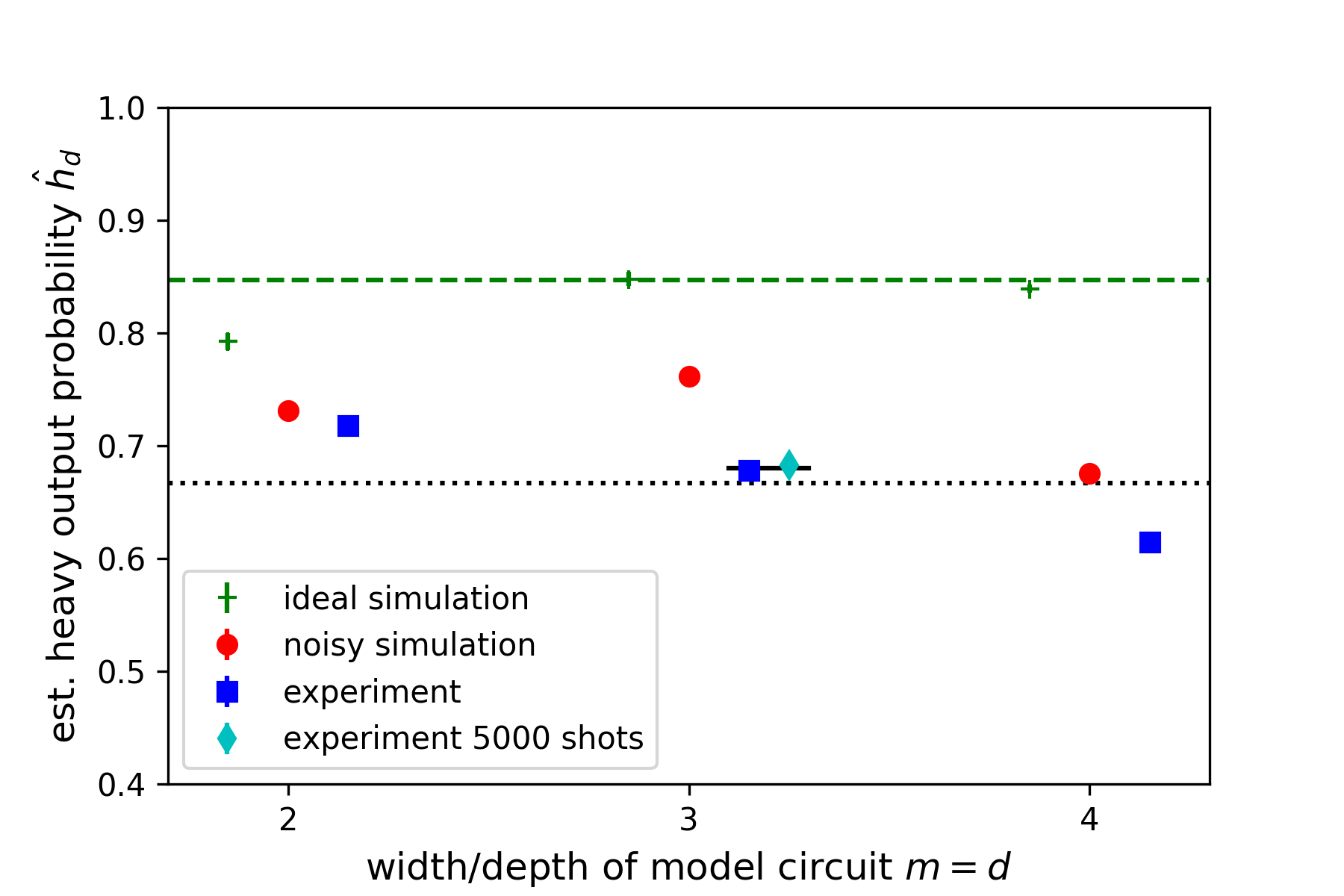}
	\caption{Experimental data for square (width = depth) quantum volume circuits using the IBM Q 20-qubit device, \textit{Tokyo}. The ideal simulation results are green plus signs. The noisy simulations, using a depolarizing noise model with average error rates from the qubits used on the device, are red circles. The experiments using 200 circuits are blue squares.  The dotted line is the threshold of 2/3 for heavy output generation, and the dashed (green) line is the asymptotic ideal heavy output probability of $\frac{1+\ln 2}{2}$ \cite{AC16}, which the ideal simulations quickly approach. In order to set a high confidence level that $h_d$ surpasses the threshold, the point at $m=d=3$ was repeated with 5000 circuits (cyan diamond).  This number of shots corresponds to a stricter threshold of 0.68 indicated by the solid line at the experimental points for $m=3$.}
	\label{fig:tokyo_data}
\end{figure}

\begin{figure}[h]
	\includegraphics[width=3.4in]{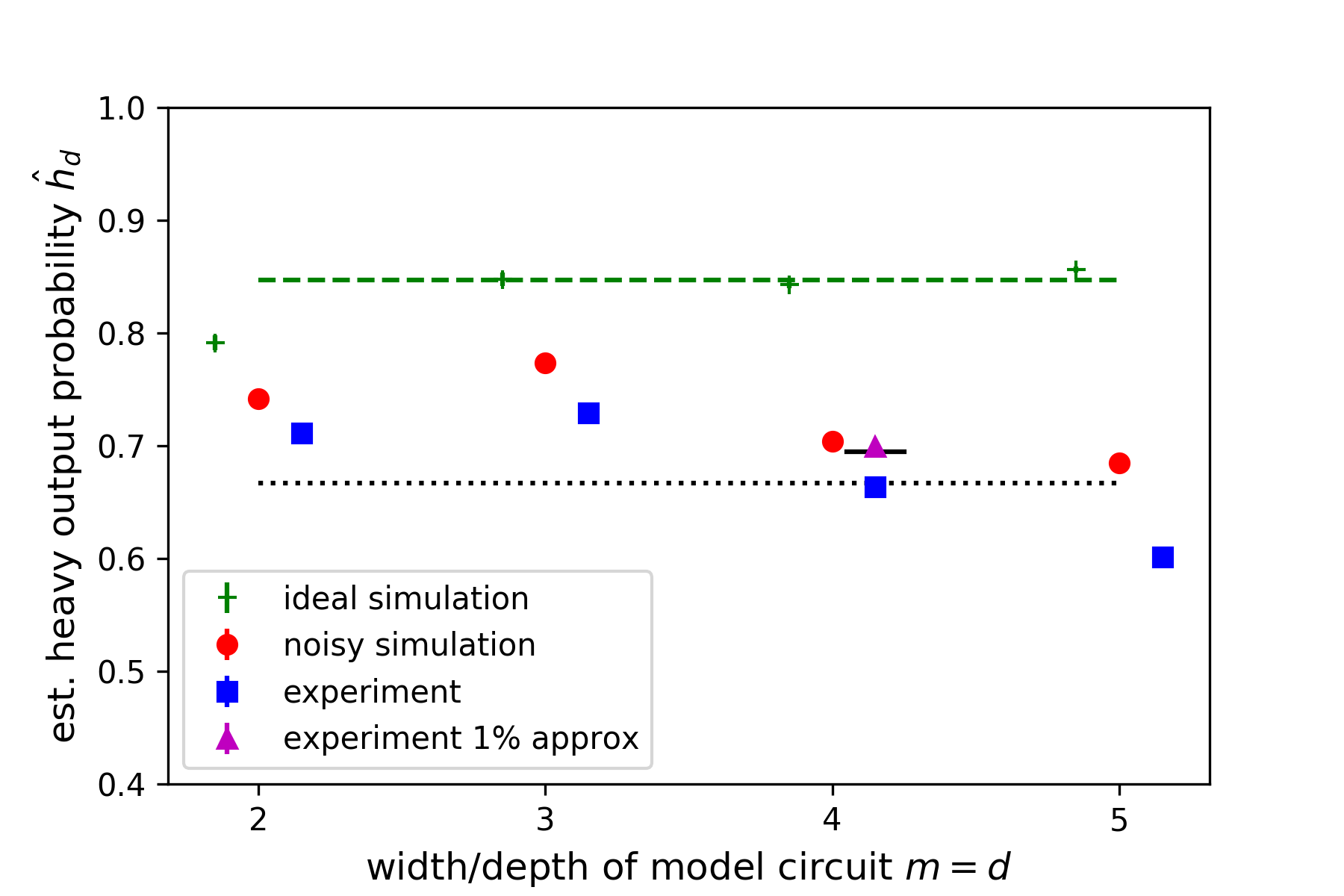}
	\caption{Experimental data for square (width = depth) quantum volume circuits using the IBM Q Johannesburg 20-qubit device. As in Figure \ref{fig:tokyo_data}, the ideal simulation results are green plus signs, the noisy simulations are red circles, and the experiments are blue squares.  Again, the dotted line is the threshold of 2/3 for heavy output generation, and the dashed (green) line is the asymptotic ideal heavy output probability.  The additional point at $m=d=4$ (magenta triangle) is not only repetition with more circuits but experimental results using optimized circuits with the KAK approximation, assuming 1\% error gates. The experiments with optimized circuits were run with 1000 circuits. The threshold for this number of circuits is 0.695 and is indicated by the solid line at $m=4$.}
	\label{fig:qs1_data}
\end{figure}

We have run quantum volume circuits on four IBM Q devices: 5-qubit \textit{Tenerife} \cite{tenerife}, 16-qubit \textit{Melbourne} \cite{melbourne}, 20-qubit \textit{Tokyo}, and 20-qubit \textit{Johannesburg}.  We generate 200 circuits for $d=m$ with $m$ in ${2,3,4}$ to determine $V_Q$.  The experimental results and comparison to simulated data for \textit{Tokyo}  and \textit{Johannesburg} are given in Figs.~\ref{fig:tokyo_data} and \ref{fig:qs1_data} respectively, whereas a summary of results across all devices is in Table \ref{table:exp_data}. We note that the noisy simulation substantially over-estimates the performance, highlighting the value of system-level metrics such as quantum volume. In order to set a high confidence level that the experimental measurements of $h_d$ surpass the threshold, we repeat the experiments for $m=2$ on \textit{Tenerife} and $m=3$ on \textit{Tokyo} with 5000 circuits.   This larger number of circuits has a strict threshold of $\hat{h}_d>0.68$ for a 97.5\% one-sided confidence interval (see Appendix \ref{app:confidence}).  From Table~\ref{table:exp_data} we see that $\log_2 V_Q=3$ for \textit{Tokyo}, $\log_2 V_Q=2$ for \textit{Tenerife}, and $\log_2 V_Q < 2$ for \textit{Melbourne}.  Additional details about the devices used here are given in Appendix~\ref{app:device}.

We also compare circuits run on \textit{Tokyo} with optimized compiling schemes.  Table \ref{table:compiled_circuits} presents $\hat{h}_d$ for $m=d=4$ found with circuits optimized both by the KAK decomposition \cite{BM03,SMB04} described in Appendix \ref{app:transpiler} and the approximate SU(4) decomposition described in Appendix~\ref{app:approximate}. The approximate decomposition takes the CX error rate as a parameter to determine acceptable approximation errors when synthesizing a circuit for an element of SU(4). We apply this decomposition assuming CX error rates of 0.01, 0.03, and 0.05 and compare the results.  We find modest increases in $\hat{h}_d$ that correspond to the reduction in the total number of CX gates in the compiled circuits: the standard Qiskit Terra transpiler \cite{QISKit} produces circuits with 28 CX gates on average, and we measure $\hat{h}_d = 0.614 (0.003)$; KAK reduces the average number of CX gates to 21 and produces $\hat{h}_d = 0.632 (0.005)$.  The approximate SU(4) circuits introduce further gains with the best result of $\hat{h}_d = 0.649 (0.005)$ achieved using circuits with a 1\% CX error approximation.

Finally, we present the outcomes of the quantum volume circuits measured on \textit{Johannesburg}.  This device has the lowest gate error rates of all the devices measured, with single qubit gate errors a factor of four smaller and two qubit gate errors nearly half than those measured on \textit{Tokyo}.  These reduced error rates suggest \textit{Johannesburg} should have the best performance of all the devices measured, and in fact we find the highest heavy output probabilities for $m=d>3$ on this device as is evident in Table \ref{table:exp_data}.  For the case $m=d=4$ the results lie just below the threshold of $\hat{h}_d = 2/3$, and optimizing the circuits with both the KAK decomposition and the approximate SU(4) with 1\% CX error yields $\hat{h}_d = 0.699 (0.001)$.

\begin{table*}
	\begin{tabular}{llllll}
		\toprule
		Circuit & \multicolumn{1}{c}{Tenerife} &  \multicolumn{1}{c}{Melbourne} & \multicolumn{1}{c}{Tokyo} & \multicolumn{1}{c}{Johannesburg} \\
		\colrule
		$m=d=2$&	0.685 (0.001)*		&	0.638 (0.006)		&	0.718 (0.006)&  0.711 (0.006)&\\
		$m=d=3$ &	0.651 (0.006)	&	0.641 (0.009)		&	 0.682 (0.002)*& 0.729 (0.007)\\
		$m=d=4$&	0.516 (0.002)	&	0.523 (0.002)		&	 0.614 (0.003)& 0.664 (0.004)\\
		$m=d=4\dagger$&		&			&	 0.649 (0.005)& 0.699 (0.001)**\\
		$m=d=5$&		&			&	 & 0.601 (0.004)\\
		\botrule
	\end{tabular}
	\caption{\label{table:exp_data}Experimentally estimated heavy output probabilities for four IBM Q devices: 5-qubit \textit{Tenerife}, 16-qubit \textit{Melbourne}, 20-qubit \textit{Tokyo}, and 20-qubit \textit{Johannesburg}, for circuits of equal width $m$ and depth $d$.  For each $m$, 200 circuits were run on every device. The experiments ($\ast$/$\ast\ast$) were repeated with (5000/1000) circuits to ensure a 97.5\% one-sided confidence interval as descriped in Appendix \ref{app:confidence}. $m=d=4\dagger$ experiments used circuits optimized with the KAK  and approximate SU(4) decompositions assuming a 1\% CX error rate.}
\end{table*}

\begin{table*}
	\begin{tabular}{cccccc}
		\toprule
		& Standard	&	KAK		&	1\% approx.		& 	3\% approx. 	&	5\% approx.\\
		\colrule
		Average \# CX Gates & 28.1	&	21.0	&	17.7	&16.1	&	15.1\\
		 Noisy Simulation &	0.676 (0.003)	&	0.687 (0.004) 	&	0.693 (0.004)	& 0.692 (0.004)	&	0.685 (0.005)	\\ 
		Experiment & 0.614 (0.003)	&	0.632 (0.005)	&	0.649 (0.005)	&	0.647 (0.005)	&	0.646 (0.005)	\\
		\botrule
\end{tabular}
\caption{Gate counts and heavy output probabilities for $m=d=4$ circuits optimized with the KAK decomposition and the approximate SU(4) decompositons assuming CX error rates of 1\%, 3\%, and 5\%.  For each width/depth, 200 circuits were run on \textit{Tokyo} and simulated using average error rates from \textit{Tokyo}. }
\label{table:compiled_circuits}
\end{table*}

To understand how the quantum volume scales in a system with limited connectivity, as gate error probabilities decrease, we consider model circuits of width $m$ on a square grid of $m$ qubits. The $m$ qubits are arranged into the largest possible square, and extra qubits are added first to a new right column and then to a new bottom row.  We approximate the achievable model circuit depth $\tilde{d}(m)$ by assuming independent stochastic errors, so that the computation fails with high probability when the model circuit volume (width times depth) satisfies
\begin{equation}\label{eq:quasianalytic}
m\tilde{d}(m) \approx \frac{1}{\epsilon_\textrm{eff}(m)}.
\end{equation}
We substitute an estimate of the mean effective error probability $\epsilon_{\textrm{eff}}(m)$ per two-qubit gate into this expression. This estimate $\epsilon_{\textrm{eff}}(m)=(a \sqrt{m}+b)\epsilon$ is proportional to the two-qubit gate error probability $\epsilon$, with a prefactor that is linear in $\sqrt{m}$. This factor fits the mean number of SWAPs necessary to bring a pair of qubits next to each other, apply the gate, and then return them to their original positions. It is twice the average shortest path length (minus one). We do a similar calculation for a loop of $m$ qubits and find $\epsilon_{\textrm{eff,loop}}(m)=(a'm+b')\epsilon$, which grows linearly with the number of qubits\footnote{For a square array, we find $a\approx 1.29$ and $b\approx -0.78$, and for a loop, we find $a'=1/2$ and $b'\approx -0.45$.}. At a given error rate $\epsilon$, we can use these expressions to estimate the quantum volume, permitting $m$ to grow as needed.

\begin{table}[h!]
	\centering
	\begin{tabular}{cccc}
		\toprule
		$\log_2 V_Q$ & All-to-All&  Square Grid & Loop \\
		&  \includegraphics[width=1.5cm]{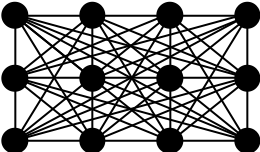}& \includegraphics[width=1.5cm]{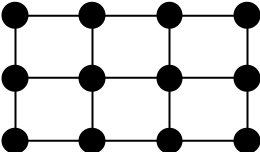} & \includegraphics[width=1.5cm]{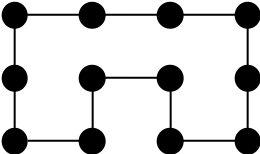}\\
		\colrule
		4	&	0.03	&	0.028		&	0.028 \\
		6	&	0.015	&	0.011		&	0.011 \\
		8	&	0.008	&	0.005		&	0.0047 \\
		12	&	0.0032	&	0.0015		&	0.0014 \\
		\botrule
	\end{tabular}
	\caption{Estimates of the maximum permissible two-qubit error needed for quantum volume $V_Q$, with $\log_2 V_Q$ given in the leftmost column, for three coupling maps: all-to-all connectivity, square grid, and loop.  The estimates are based on simulations using a depolarizing noise model with two-qubit error $\epsilon$ as given, single-qubit error equal to $\epsilon/10$, and  perfect measurements.  }
\label{table:connectivity_sim}
\end{table}		
	\begin{table}[h!]
		\centering
		\begin{tabular}{cccc}
		\toprule
		$\log_2 V_Q$&  0\%  meas. error &  1\% meas. error & 5\% meas. error \\
		\colrule
		4	&	0.028		&	0.026 	&	0.020\\
		6	&	0.011		&	0.010 	&	0.007\\
		8	&	0.005		&	0.0045	&	 0.0023\\
		12	&	0.0015		&	0.00125	&	 0.0002\\
		\botrule
	\end{tabular}
	\caption{A comparison of the maximum permissible two-qubit error rate for $\log_2 V_Q$ of 4, 6, 8, and 12 for three values of the measurement error: 0\%, 1\%, and 5\%. These simulations all use a square grid coupling map;  the 0\% measurement error column is identical to the square grid column of Table \ref{table:connectivity_sim}.  }
	\label{table:connectivity_sim_meas}
\end{table}

To validate these estimates, we consider the influence of connectivity on quantum volume by simulating three coupling graphs for up to 12 qubits: all-to-all connectivity, square grid, and loop. We estimate the two-qubit gate error $\epsilon$ required for each coupling graph to obtain a $\log_2 V_Q$ of 4, 6, 8, and 12, assuming the single-qubit gate error is equal to $\epsilon/10$ (Table \ref{table:connectivity_sim}).   We run these simulations with no measurement error for all graphs, and for measurement errors of 0\%, 1\%, and 5\% for the square grid (Table \ref{table:connectivity_sim_meas}). The values for $\epsilon$ here correspond to 200 simulated circuits with a heavy output probability of $\hat{h}_d=0.67\pm0.05$.  

\begin{figure}[h]
	\includegraphics[width=3.2in]{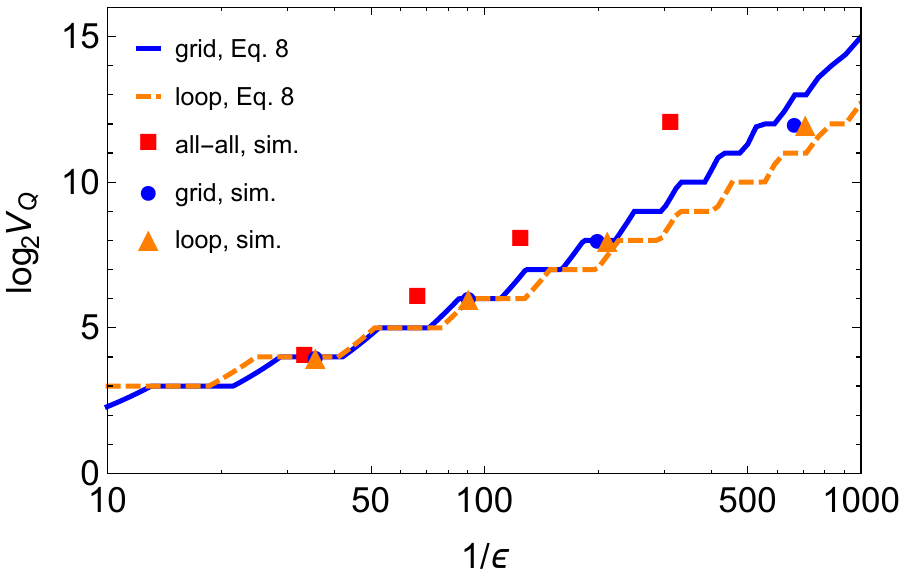}
	\caption{The quantum volume increases as a function of inverse gate error $1/\epsilon$. This plot shows numerical simulation results from the top half of Table~\ref{table:connectivity_sim} together with estimates using the expression in Eq.~\ref{eq:quasianalytic} for grid and loop connectivities.\label{fig:numerics}}
\end{figure}

It is clear from Table \ref{table:connectivity_sim} that all-to-all connectivity provides an advantage over the less-connected graph; $\log_2 V_Q$ of 12 is achievable with twice the two-qubit error rate (0.0032) of the square grid (0.0015) and the 12-qubit loop (0.0014). At the same time, there is little difference between the required two-qubit error rate for the square grid versus the loop graphs; the error rate for the loop is less than 7\% lower than that of the square grid for the 12-qubit case. This relatively small difference is due to the small total number of qubits, since there is a significant asymptotic difference between loop and grid layouts. However, the difference may increase, even at small sizes, when using an optimal transpiler.  All circuits for the simulations in Tables \ref{table:connectivity_sim} and \ref{table:connectivity_sim_meas} were compiled using the standard Qiskit Terra transpiler. Quantum volume estimates computed from Eq.~\ref{eq:quasianalytic} are consistent with these depolarizing noise simulations at error probabilities down to $\epsilon\approx 10^{-3}$, as shown in Fig.~\ref{fig:numerics}.

These simulations give an indication of how quantum volume measurements might look on different quantum computing architectures.  Trapped ions, for instance, will benefit from having all-to-all connectivity. Typical trapped-ion systems have both two-qubit gate errors and measurement errors less than 0.01, which based on Table \ref{table:connectivity_sim} should be sufficient to achieve $\log_2 V_Q=6$ if not higher.  Recently, trapped-ion experiments have demonstrated two-qubit gates with errors of 0.001 \cite{Ballance2016}, indicating higher quantum volumes should be possible. However, multi-qubit experiments are susceptible to larger error rates than isolated two-qubit gates, due to correlated errors across many ions \cite{Monz2011}. A measurement of quantum volume would give a reliable validation of multi-qubit trapped-ion systems.  Similarly, we can infer that for superconducting devices, coupling maps with more connectivity should produce higher quantum volume, but only if additional coupling does not also introduce larger errors.

{\bf Conclusion:} In this paper we expand on a previously presented metric, the quantum volume~\cite{qvolume,Moll18}, and show both a concrete specification and a method for benchmarking noisy intermediate-scale quantum devices. This metric takes into account all relevant hardware parameters. This includes the performance parameters (coherence, calibration errors, crosstalk, spectator errors, gate fidelity, measurement fidelity, initialization fidelity) as well as the design parameters such as connectivity and gate set. It also includes the software behind the circuit optimization. Additionally, the quantum volume is architecture-independent, and can be applied to any system that is capable of running quantum circuits.  We implement this metric on several IBM Q devices, and find that we can successfully implement model circuits on up to $\log_2 V_Q=4$ qubits, which corresponds to a quantum volume as high as $V_Q=16$.  We conjecture that systems with higher connectivity will have higher quantum volume given otherwise similar performance parameters.

From numerical simulations for a given connectivity, we find that there are two possible paths for increasing the quantum volume. Although all operations must improve to increase the quantum volume, the first path is to prioritize improving the gate fidelity above other operations, such as measurement and initialization. This sets the roadmap for device performance to focus on the errors that limit gate performance, such as coherence and calibration errors.  The second path stems from the observation that, for these devices and this metric, circuit optimization is becoming important. We implemented various circuit optimization passes (far from optimal) and showed a measurable change in the experimental performance.   In particular, we introduced an approximate method for NISQ devices, and used it to show experimental improvements. 

We encourage the adoption of quantum volume as a primary performance metric, which we believe will allow the field to work together and focus efforts on the important factors to develop improved NISQ devices. To this end, we have released a library for measuring quantum volume as an open-source component of Qiskit Ignis \cite{Qiskit}.

\begin{acknowledgments}
The authors acknowledge support from ARO under Contract No.\ W911NF-14-1-0124 and thank Sergey Bravyi, John~A.~Smolin, and Christopher~J.~Wood for informative discussions. We thank Antonio C\'orcoles, Abigail Cross, John Gunnels, David McKay, Travis Scholten, and Ted Yoder for valuable comments on the manuscript. We are grateful to the IBM Q team for their contributions to the systems and devices used in this work.
\end{acknowledgments}

\appendix

\section{Qiskit transpiler passes}\label{app:transpiler}
 
Model circuits must be rewritten to use the gate set of the target system, while attempting to minimize any additional overhead that might result from the translation. The IBM Q systems used in this paper accept quantum circuits expressed by products of controlled-NOT (CNOT) gates and single-qubit gates \cite{openqasm}. The single-qubit gates are defined by
 \begin{align}
u_1(\lambda) & = \mathrm{diag}(1,e^{i\lambda}) \\
u_2(\phi, \lambda) & = R_z(\phi+\pi/2)R_x(\pi/2)R_z(\lambda-\pi/2) \\
u_3(\theta, \phi, \lambda) & = R_z(\phi+3\pi)R_x(\pi/2)R_z(\theta+\pi)R_x(\pi/2)R_z(\lambda)
 \end{align}
 where $R_P(\theta)=\mathrm{exp}(-i\theta P/2)$ for a Pauli matrix $P\in \{X, Y, Z\}$. The available CNOT gates for a particular system are given in the form of a qubit connectivity graph $G=(V, E)$. Each vertex of $G$ represents a qubit and each (directed) edge represents a pair of qubits that can be coupled by gates.
 
We generate input model circuits by sampling and expanding each SU(4) gate to CNOT and single-qubit gates using the KAK decomposition \cite{BM03,SMB04} implemented in Qiskit Terra (see also Appendix~\ref{app:approximate}). Each input circuit is then mapped to the target system and optimized using a sequence of circuit rewriting passes that are implemented in Qiskit Terra. These passes are named unrolling, CNOT reorientation, CNOT cancellation, single-qubit optimization, and swap mapping. All of the passes can be applied multiple times, but some passes, such as CNOT reorientation, have requirements that are ensured by other passes, such as swap mapping.

The unrolling pass is essentially a macro expansion that descends into each gate's hierarchical definition and rewrites that gate in terms of lower-level gates. In the setting of rewriting model circuits, the lower-level gate set is always the IBM Q gate set. For example, a Hadamard ($H$) gate is defined as $u_2(0,\pi)$ in the Qiskit Terra gate library, which is in the IBM Q gate set, and a SWAP gate is defined as $\CNOT_{a,b}\CNOT_{b,a}\CNOT_{a,b}$.

The CNOT reorientation pass examines each CNOT gate in the circuit and applies the identity
\begin{equation}
\CNOT_{c,t}=(H\otimes H)\CNOT_{t,c}(H\otimes H)
\end{equation} 
if $(t,c)$ is a directed edge of $G$ but $(c,t)$ is not. The pass fails if neither $(c, t)$ nor $(t, c)$ are edges of $G$.

The CNOT cancellation pass collects sequences $\CNOT_{c,t}^m$ of CNOT gates with the same control and target qubits, and replaces them by $\CNOT_{c,t}$ if $m$ is odd or removes them from the circuit if $m$ is even.

The single-qubit optimization pass collects sequences of single-qubit gates on the same qubit and replaces each sequence by at most one single-qubit gate. Furthermore, the replacement is chosen in an attempt to minimize the number of physical pulses used to implement the gate; $u_1$ uses zero pulses, $u_2$ uses one pulse, and $u_3$ uses two pulses. The algorithm composes the gates in sequence, rewriting each composed pair of gates as a new gate according to a handful of rewriting rules that follow from the definitions.

The swap mapping pass is the most involved of the fundamental passes within Qiskit Terra. This pass first partitions the input circuit into a sequence of layers such that each layer consists of gates that act on disjoint sets of qubits. The algorithm then acts layer by layer. For simplicity we will ignore single-qubit gates in the following discussion. Consider the gate $U=U_1U_2\dots U_m$ applied in a particular layer, where $U_1, \dots, U_m$ are pairwise disjoint two-qubit gates that may act on remote pairs qubits. When the mapping pass acts on this layer, it computes a quantum circuit $U'$ with the following properties:
\begin{enumerate}
	\item $U'$ consists of nearest-neighbor gates with respect to the connectivity graph $G=(V,E)$
	\item $U'=WU$ where $W$ is some permutation of the $n=|V|$ qubits
	\item $U'$ has small depth, which the algorithm tries to minimize subject to the first two conditions
\end{enumerate}
The algorithm to compute $U'$ consists of a sequence of rounds, each of which increases the depth of $U'$ by one. At the beginning of a round, the algorithm applies all gates $U_j$ that are nearest-neighbors and removes them from $U$. The rest of the round performs a greedy (randomized) optimization over swap gates to choose a depth-one swap circuit that brings pairs of qubits coupled by gates as close as possible.

The passes are applied in the following order for our standard compilation:
\begin{enumerate}
	\item Unrolling pass
	\item Swap mapping pass
	\item Unrolling pass (to expand SWAP gates)
	\item CNOT reorientation pass
	\item CNOT cancellation pass
	\item Unrolling pass (to expand Hadamard gates)
	\item Single-qubit optimization pass
\end{enumerate}
In our study of optimized model circuits, we apply the following optimization passes after the standard set of passes:
\begin{enumerate}
	\item Two-qubit block collection pass
	\item Two-qubit block optimization pass
\end{enumerate}
The two-qubit block collection pass is an analysis pass that traverses the circuit's gates in topologically sorted order. Starting at each newly-discovered CNOT gate, the pass explores that gate's predecessors and ancestors to collect the largest block of previously unseen and contiguous gates acting on the control and target qubits. The pass continues in this manner and returns a collection of disjoint blocks. The two-qubit block optimization pass computes the unitary operation for each block, synthesizes a new sub-circuit (either exactly, using the KAK decomposition \cite{BM03, SMB04}, or approximately; see Appendix~\ref{app:approximate}), and replaces the block.

To further reduce the number of SWAP gates, we considered an optimization called the Local Ordering Circuit Optimization (LOCO), that permutes qubits such that those interacting via CNOT gates are as nearest-neighbor as possible in the circuit representation; the circuit is optimized for a linear nearest-neighbor  topology.  This method employs a weighted-variant of reverse Cuthill-Mckee ordering \cite{Cuthill69, George81} to reorder the sparse matrix $A_{ij}$, with non-zero elements counting the number of CNOT gate operations between qubits $i$ and $j$ in the circuit, so that its bandwidth is minimized. The matrix is symmetric as we do not consider the direction of the CNOTs.  This reordering is efficient, having a runtime that is linear in the number of nonzero matrix elements \cite{Chan80}.  To properly account for multiple CNOT interactions between qubits, the LOCO algorithm uses a weighted heuristic when reordering, that favors optimizing pairs of qubits with the largest number of repeated interactions over those with fewer gates between them. Input circuits whose bandwidth was reduced by LOCO were replaced with their optimized counterparts. Although this optimization did not lead to significant improvements for heavy output generation using small numbers of qubits, we expect SWAP optimizations such as these to further improve results for larger circuits mapped onto devices with limited connectivity.

 \section{Approximate compiling}\label{app:approximate}
 We can always decompose~\cite{kraus_optimal_2001,*khaneja_time_2001} an arbitrary two-qubit unitary in the form
\begin{gather}
U=K_1U_d(\alpha, \beta, \gamma)K_2 ,
\end{gather}
where $K_i=K_i^l\otimes K_i^r$ are products of single-qubit unitaries $K_i^{l,r}$, the two-qubit component is represented in terms of the \textit{information content} $(\alpha, \beta, \gamma)$ as
\begin{gather}
U_d(\alpha, \beta, \gamma) = \exp[i ( \alpha \sigma_x\otimes \sigma_x + \beta \sigma_y\otimes \sigma_y + \gamma \sigma_z\otimes \sigma_z)] , 
\end{gather}
and we can always restrict to the Weyl chamber $\pi/4\geq\alpha\geq\beta\geq|\gamma|$.  
Let $U\sim V$ denote equivalence between $U$ and $V$ under local operations, implying equality of the information content of $U$ and $V$.

We can calculate a trace  of the product of two  $U_i=U_d(\alpha_i,\beta_i,\gamma_i)$ as 
\begin{align}
\begin{split}
\tr(U_c^\dagger U^{\phantom{\dagger}}_t)&= 4\cos(\Delta_\alpha)\cos(\Delta_\beta)\cos(\Delta_\gamma)\\
&\quad-4 i \sin(\Delta_\alpha)\sin(\Delta_\beta)\sin(\Delta_\gamma),
\end{split}
\end{align}
where 
\begin{subequations}
\begin{align}
\Delta_\alpha&=\alpha_c-\alpha_t,\\
\Delta_\beta&=\beta_c-\beta_t,\\
\Delta_\gamma&=\gamma_c-\gamma_t.
\end{align}
\end{subequations}
From this trace we may easily determine the average gate fidelity~\cite{horodecki_general_1999}
\begin{gather}
F_\text{avg}(U_c,U_t) =\frac{4+\bigl|\tr(U_c^\dagger U^{\phantom{\dagger}}_t)\bigr|^2}{20}\label{eq:fid}
\end{gather}
and these expressions give also the maximal fidelity between arbitrary unitaries $U_{c,t}\in\mathrm{SU}(4)$ after optimizing over local pre- and post-rotations~\cite{watts_optimizing_2015}
\begin{gather}
\max_{K_1^l, K_1^r, K_2^l, K_2^r} F_\text{avg}\bigl[(K_1^l\otimes K_1^r)U_c(K_2^l\otimes K_2^r),U_t\bigr] .
\end{gather}

We are interested in decompositions of a target unitary $U_t\in \mathrm{SU}(4)$ with the minimal number of applications of a fixed `basis' gate $U_b$.
It is obvious that with zero applications of the basis we can construct only non-entangling target unitaries $U_t\sim U_d(0,0,0)$, and with one application of the basis we can construct only target unitaries which are equivalent to the basis $U_t\sim U_d(\alpha_b, \beta_b, \gamma_b)$.
For $U_b\sim \CNOT\sim U_d(\pi/4,0,0)$ it is well-known~\cite{vatan_optimal_2004, *vidal_universal_2004} that 3 applications of the basis is sufficient to cover all of SU(4). 
Zhang et al.~\cite{zhang_conditions_2005} give decompositions using a more general `super controlled' basis $U_b\sim U_d(\pi/4,\beta_b,0)$, for any $\beta_b$, both an expansion with 3 applications of $U_b$  to decompose an arbitrary $U_t\sim U_d(\alpha_t, \beta_t, \gamma_t)$ and also an expansion using two applications of $U_b$ for a restricted target unitary $U_t\sim U_d(\alpha_t,\beta_t,0)$, $\gamma_t=0$ for any $\alpha_t$, $\beta_t$. 

The above expansions are \emph{exact} so that the constructed unitary $U_c$ satisfies
\begin{gather}
F_\text{avg}(U_t,U_c) = 1 , 
\end{gather}
but we can use eq.~(\ref{eq:fid}) to find the average gate fidelity due to approximating general $U_t$ by fewer applications of the basis gate than is necessary for exact expansion. 
With zero applications of arbitrary $U_b$ we have:
\begin{subequations}\label{eqs:expand}
\begin{align}
U_c^{(0)} &= K_{t,1}K_{t,2},\\
F^{(0)}_\text{avg}&=\Bigl[1+4\cos^2(\alpha_t)\cos^2(\beta_t)\cos^2(\gamma_t)\nonumber\\&\qquad+4\sin^2(\alpha_t)\sin^2(\beta_t)\sin^2(\gamma_t)\Bigr]/5, 
\end{align}
which is optimal. With one application of arbitrary $U_b$ we have:
\begin{align}
U_c^{(1)} &=K_{t,1}U_d(\alpha_b, \beta_b, \gamma_b)K_{t,2},\\
F^{(1)}_\text{avg}&=\Bigl[1+4\cos^2(\Delta_\alpha)\cos^2(\Delta_\beta)\cos^2(\Delta_\gamma)\nonumber\\&\qquad+4\sin^2(\Delta_\alpha)\sin^2(\Delta_\beta)\sin^2(\Delta_\gamma)\Bigr]/5, 
\end{align}
which is optimal.
With two applications of super controlled $U_b\sim U_d(\pi/4,\beta_b,0)$ we have:
\begin{align}
U_c^{(2)} &= K_{t,1}U_d(\alpha_t, \beta_t, 0)K_{t,2},\\
F^{(2)}_\text{avg}&=\Bigl[1+4\cos^2(\gamma_t)\Bigr]/5, \label{eq:twobasisfid}
\end{align}
which is optimal for $U_b\sim\CNOT\sim U_d(\pi/4,0,0)$ or $U_b\sim\mathrm{DCNOT}\sim U_d(\pi/4,\pi/4,0)$.
For completeness, with 3 applications of super controlled $U_b$ there is no need to approximate and we have:
\begin{align}
U_c^{(3)} &= K_{t,1}U_d(\alpha_t, \beta_t, \gamma_t)K_{t,2} = U_t,\\
F^{(3)}_\text{avg}&=1, 
\end{align}
which is clearly optimal.
\end{subequations}

\begin{figure}[tbp]
\includegraphics[width=\columnwidth]{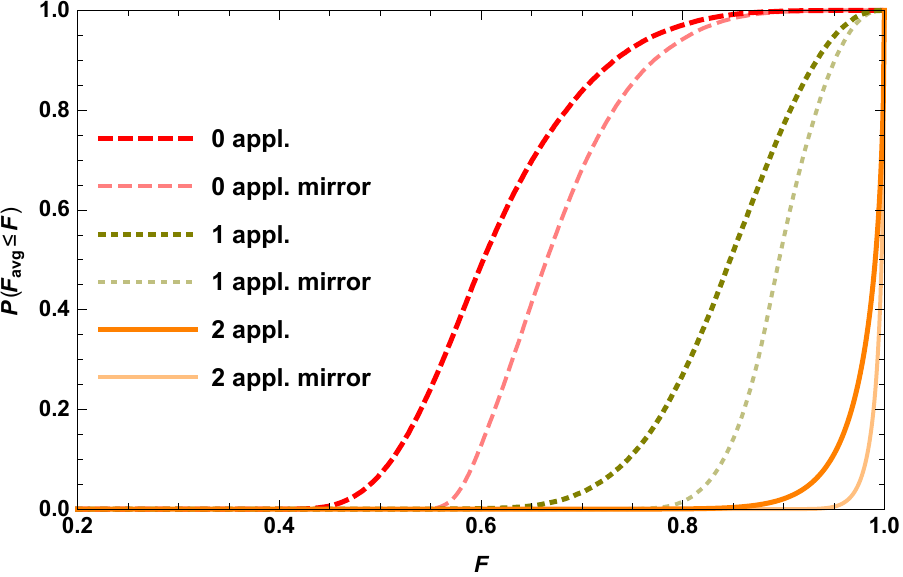}
\caption{\label{fig:approxfid}(Color online) Average gate fidelity for random target gates in the Haar measure, for approximations using  zero, one, or two applications of a 2-qubit super controlled basis gate, with and without freedom to mirror. These approximations are optimal for the case that the basis gate is equivalent to CNOT.}
\end{figure}

There can be an additional freedom when expanding a two-qubit gate:  in many cases it does not matter whether we implement $U_t$ or $U_{tm}=U_t\cdot\mathrm{SWAP}$ since the latter differs merely by permutation of the output qubit labels. We call it the \textit{mirror gate} of $U_t$ and its expansion is easily related to $U_t$:
\begin{gather}
U_{tm}\sim U_d\bigl(\pi/4-|\gamma_t|, \pi/4-\beta_t,\sign(\gamma_t)(\alpha_t-\pi/4)\bigr) ,
\end{gather}
making use of the sign function defined as $\sign(x)=-1$ for $x<0$ and $\sign(x)=1$ for $x\geq0$. We can extend eqs.~(\ref{eqs:expand}) to give $i$-gate expansions of $U_{tm}$, $U_c^{(im)}$ with fidelities $F^{(im)}_\text{avg}$, defined by choosing to expand whichever of $U_t$ and $U_{tm}$ gives the better fidelity. For example, the 2-gate expansion has
\begin{gather}\label{eq:twobasisfidm}
F^{(2m)}_\text{avg}=\Bigl[1+4\cos^2\Bigl(\min\bigl[|\gamma_t|,|\alpha_t-\pi/4|\bigr]\Bigr)\Bigr]/5 .
\end{gather}
Because of the mirroring action within the Weyl chamber, the expansion of the mirrored gate has best fidelity exactly when the expansion of the unmirrored gate has worst fidelity, and vice versa. In addition to improving $F_\text{avg}$, the freedom to combine a SWAP operation may also allow reduction in the number of inserted SWAP gates during a `swap mapping pass' as described in Appendix~\ref{app:transpiler}.

It is interesting to investigate the expected infidelity of each of the approximate expansions of $U_t$, averaged over $U_t$ uniformly distributed within $\mathrm{SU}(4)$ in the Haar measure on the Weyl chamber~\cite{watts_metric_2013,*musz_unitary_2013}
\begin{align}
M(\alpha,\beta,\gamma)&= \frac{24}{\pi}\Bigl[\cos(4\alpha)\cos(8\beta)+\cos(4\beta)\cos(8\gamma)\nonumber\\
&\quad + \cos(4\gamma)\cos(8\alpha)-\cos(8\alpha)\cos(4\beta)\nonumber\\
&\quad - \cos(8\beta)\cos(4\gamma)-\cos(8\gamma)\cos(4\alpha)\Bigr] ,
\end{align}
allowing calculating the distribution of fidelities of the 2-basis gate approximation of eq.~(\ref{eq:twobasisfid}) for a random element of $\mathrm{SU}(4)$
\begin{multline}
P(F_\text{avg}^{(2)}<F) = \cos^4(2z)\Bigl[\bigl(4z-\pi\bigr)\bigl(\cos(4z)-2\bigr)\\
-3\sin(4z)\bigr]/\pi ,
\end{multline}
where $z$ is defined by
\begin{align}
\cos(z) = \frac{\sqrt{5F-1}}{2},
\end{align}
for $F>3/5$, and
\begin{gather}
P(F_\text{avg}^{(2)}<F) = 0
\end{gather}
for $F\leq3/5$. Similarly, for the mirrored version eq.~(\ref{eq:twobasisfidm})
\begin{multline}
P(F_\text{avg}^{(2m)}<F) =\cos (4 z) \Bigl[\bigl(8 z-\pi\bigr)\bigl(\cos (8 z)-2\bigr)\\-3 \sin (8 z)\Bigr]/\pi  ,
\end{multline}
for $z<\pi/8$, $F>0.88$, and 
\begin{gather}
P(F_\text{avg}^{(2m)}<F) = 0
\end{gather}
for $z\geq\pi/8$.

The 2-basis gate approximations perform surprisingly well, with the median fidelities $F^{(2)}_\text{avg}=0.99$, $F^{(2m)}_\text{avg}=0.997$ comparing favorably to the typical 2-qubit gate fidelities for current quantum devices. The full distribution of fidelities for the zero-, one-, and two-gate approximations are plotted in Fig.~\ref{fig:approxfid}, where the zero- and one-gate distributions are determined by random sampling.

By comparing $F_\text{avg}^{(i)}$ for all $i$ we can choose the best approximation for any given $U_t$. Specifically, if the basis gate  $U_b$ may be implemented with average gate fidelity $F_b$ we can estimate the overall fidelity by multiplying the fidelity due to approximation with the fidelity due to the number of applications of $U_b$, and choose the expansion with the highest overall fidelity
\begin{subequations}\label{eq:argmin}
\begin{align}
F_\text{best}&=\max_i F_\text{avg}^{(i)} (F_b)^i, \label{eq:argminno}\\
F_\text{best}^{(m)}&=\max_i F_\text{avg}^{(im)} (F_b)^i. \label{eq:argminyes}
\end{align}
\end{subequations}

\begin{figure}[tbp]
\includegraphics[width=\columnwidth]{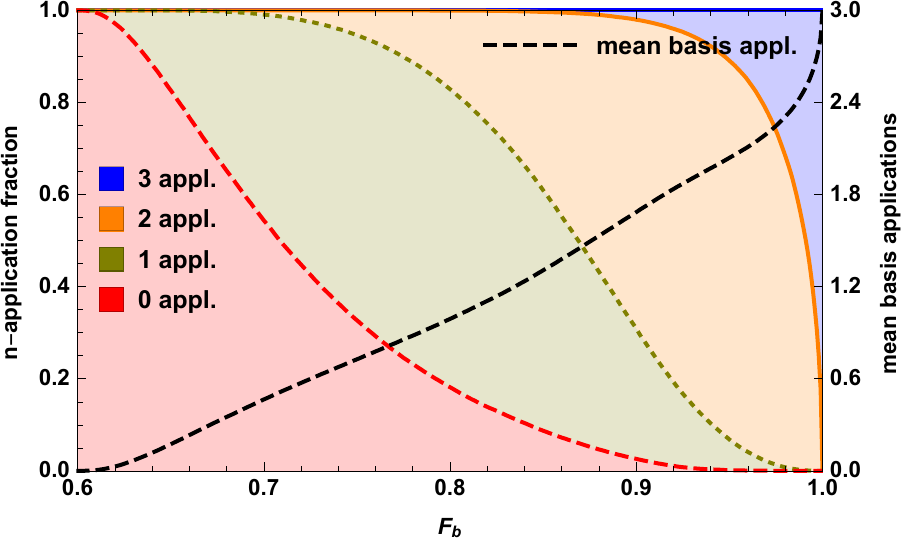}
\includegraphics[width=\columnwidth]{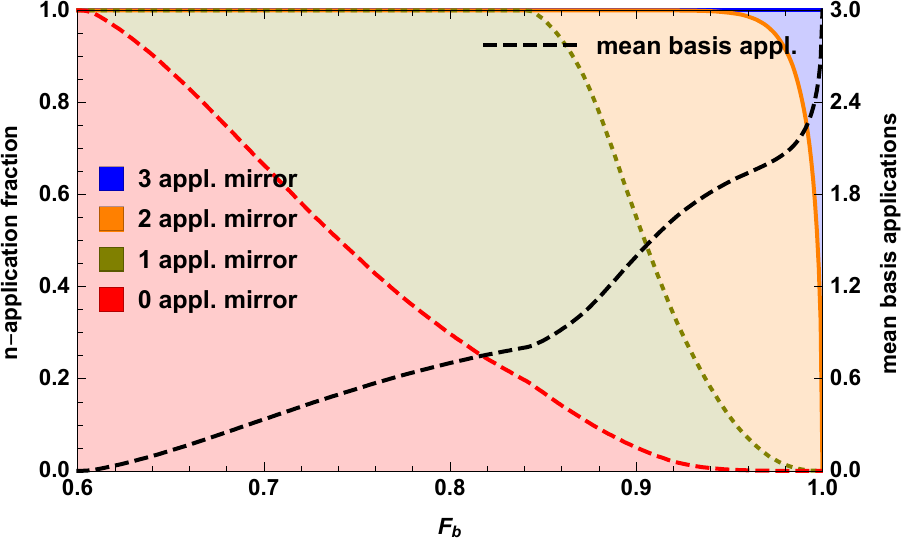}
\caption{\label{fig:ngate}(Color online) Number of basis gate applications used for approximating chosen for randomly chosen target gates in the Haar measure, choosing the approximation according to eq.~(\ref{eq:argmin}) as a function of the basis gate fidelity $F_b$. Fraction of cases with each number of applications is shown by shading (left axis) and the mean number of basis applications is shown by the dashed line (right axis). (a)~without mirroring, as in eq.~(\ref{eq:argminno}), (b)~with mirroring, as in eq.~(\ref{eq:argminyes})}
\end{figure}

\begin{figure}[tbp]
\includegraphics[width=\columnwidth]{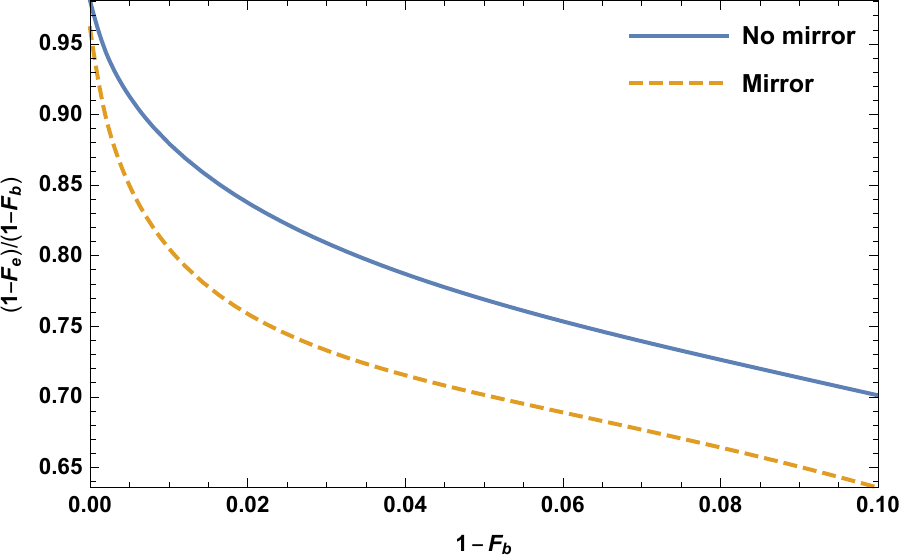}
\caption{\label{fig:fidratio}(Color online) Effective infidelity ratio as a function of basis gate infidelity, with and without freedom to mirror. }
\end{figure}

The statistics of the number of basis gate applications for a randomly-generated ensemble of target gates are shown in Fig.~\ref{fig:ngate}. With a fairly noisy basis gate $F_b=0.97$ and no mirroring, the best expansion by this method has 3 applications of the basis for $22\%$, two applications for $76\%$,  one application for $2\%$, and zero applications for $<0.1\%$ of targets, thus an average of 2.2 basis gate applications. With the freedom to mirror, three applications for $3\%$, two applications for $93\%$, one application for $4\%$,  and zero applications for $<0.1\%$ of targets, thus a mean of 2.0 basis gate applications. The resulting fidelity can be quoted as an `effective fidelity' $F_e$ equal to the cube root of the mean of $F_\text{best}$, which we can interpret as the equivalent basis gate fidelity if we were to use only exact 3-gate expansions of random targets. We show in Fig.~\ref{fig:fidratio} the ratio of the effective infidelity $1-F_e$ to the basis gate infidelity $1-F_b$, giving the factor by which the use of approximate expansions improves effective gate performance. For $F_b=0.97$ we get $F_e=0.976$, $F^{(m)}_e=0.978$, reducing the infidelity by factors of 0.82 and 0.74 respectively.

For the volume measurements described in the main text, Table~\ref{table:compiled_circuits}, we implemented the approximate two-qubit block optimization compilation pass without mirroring, assuming fixed $1-F_b$ of  $1\%$, $3\%$ or $5\%$. Because the 4 qubits chosen for the \textit{Johannesburg} have a linear nearest-neighbor topology, we were able to implement a special-case optimization that replaces some gates by the corresponding mirrored gate in order to minimize the number of inserted SWAP gates for this topology. Using measured CNOT fidelities for each of the qubit pairs, implementing the mirror expansions, and combining the mirror choice with a swap-mapping pass for general topologies should allow future compiler-driven improvements in quantum volume.

\section{Confidence intervals for the heavy probability}\label{app:confidence}

To be confident with a finite number of trials that the heavy probability $h_d$ exceeds $2/3$, we should set stricter threshold $t>2/3$, requiring the estimated heavy probability $\hat{h}_d>t$ to claim success. This is a hypothesis test with null hypothesis $H_0: h_d=2/3$ and alternative hypothesis $H_1: h_d>2/3$. Drawing $n_c$ random model circuits of given width and depth, and executing each circuit $n_s$ times gives a total of $n_c n_s$ experiment outcomes, each of which is to be checked against simulation of the corresponding circuit to determine a count $n_h$ of heavy outcomes. We estimate $h_d$ in the natural way by the heavy fraction over these outcomes
\begin{gather}
\hat{h}_d =  \frac{n_h}{n_c n_s} .
\end{gather}

For the purposes of making a conservative bound on the spread of $\hat{h}_d$ we analyze using the worst-case distribution where the heavy probability conditioned on each circuit is either zero or one. Thus, executing each circuit multiple times $n_s>1$ (as is typically convenient to avoid reconfiguring experimental settings and allow recycling of simulation results) will generally narrow the observed fluctuations in $\hat{h}_d$ but, for fear of systematic errors we do not allow this to alter the threshold $t$. Under this worst-case assumption,  $n_h/n_s$ is  binomial distributed with parameter $n_c$.

While it would be straightforward to calculate numerically confidence intervals directly from the binomial distribution, because the interesting range of $\hat{h}_d$ is close to $2/3$ where a normal approximation is valid, we instead require a minimum of $n_c=100$ circuits and make a normal approximation to the binomial, and write the requirements for claiming success at a given width and depth 
\begin{gather}
n_c\geq100\\
\frac{n_h-z\sqrt{n_h(n_s-\frac{n_h}{n_c})}}{n_c n_s}>\frac{2}{3} ,
\end{gather}
where we set $z=2$ for a $97.5\%$ `2-sigma' one-sided confidence interval. For example, to claim success with $n_c=5000$ model circuits, the observed heavy fraction must exceed the threshold $t=0.68$.
 
\section{Device parameters}\label{app:device}

\begin{figure*}
	\begin{align*}
	(a) &~\includegraphics[width = 2in]{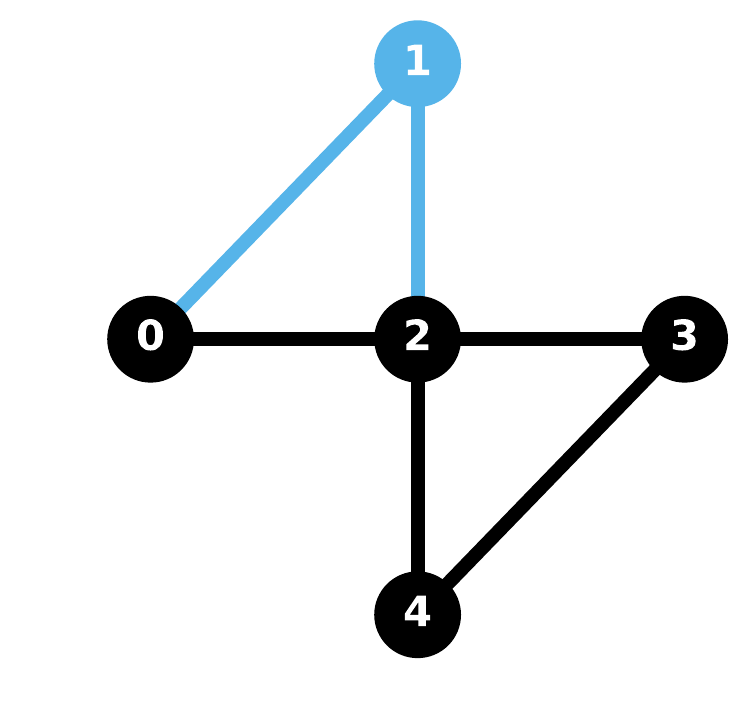}\hspace{0.5in}
	 (b)~\includegraphics[width = 4in]{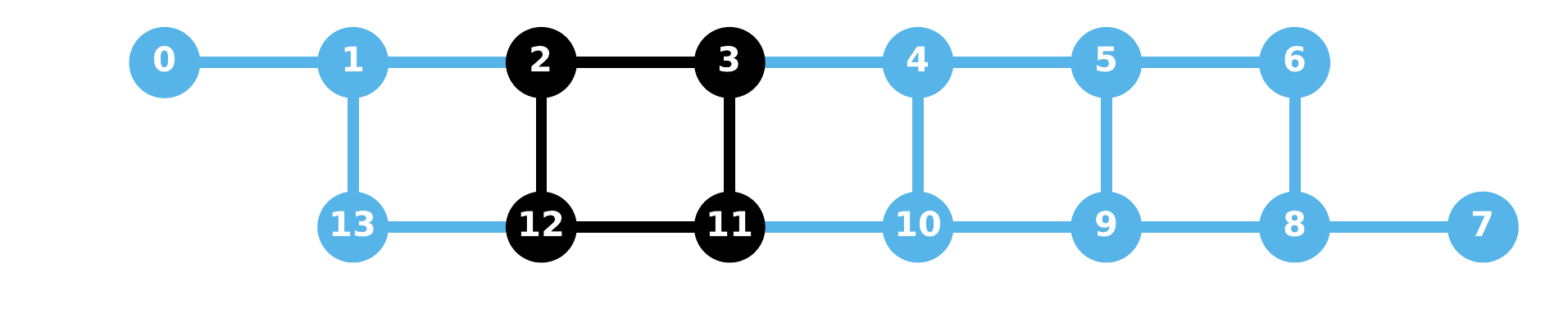}\\
	 (c)& ~\includegraphics[width = 2.5in]{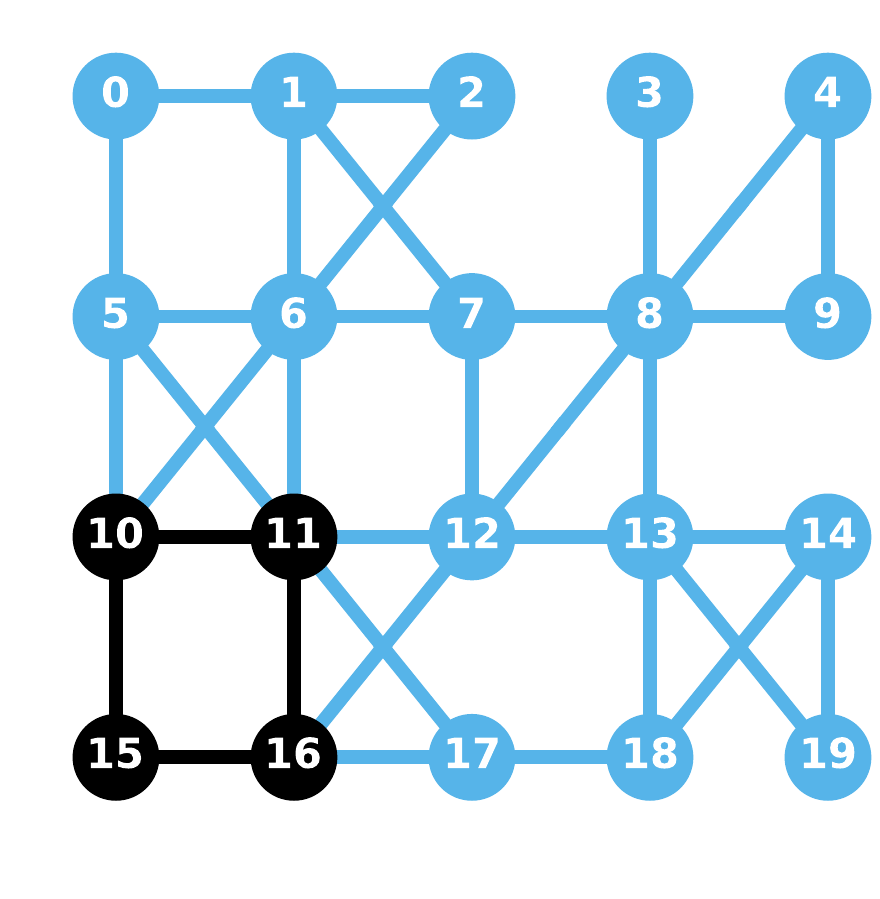}\hspace{0.5in}
	 (d) ~\includegraphics[width = 2.5in]{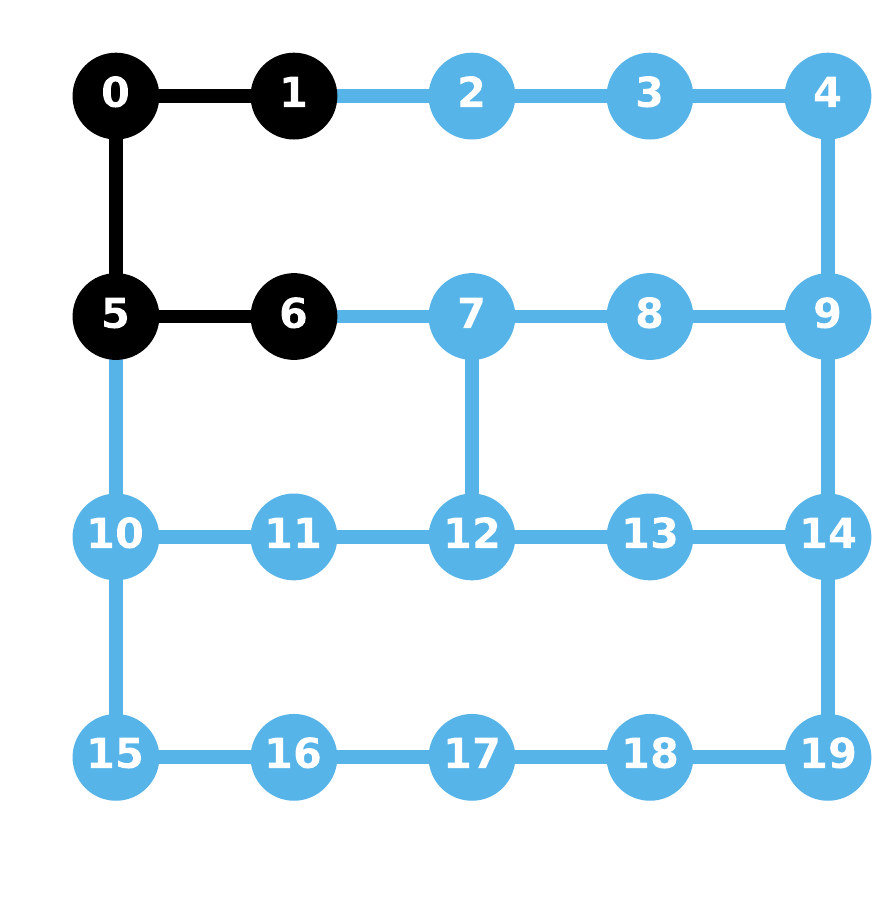}
	\end{align*}
	\caption{Device diagrams used for the experimental data in Table \ref{table:exp_data}: (a) \textit{Tenerife}, (b) \textit{Melbourne}, (c) \textit{Tokyo}, and (d) \textit{Johannesburg}.  The highlighted qubits are those selected for the experiments discussed here.  CX gates are available between pairs of qubits connected by a highlighted line.}
	\label{fig:deviceimages}
\end{figure*}

We measured the quantum volume of four IBM Q devices: 5-qubit \textit{Tenerife}, 16-qubit \textit{Melbourne}, and 20-qubit \textit{Tokyo}, and 20-qubit \textit{Johannesburg}. The device connectivities are shown in Fig. \ref{fig:deviceimages}, with the four qubits from each device that were used for the $m=d=4$ experiments highlighted in grey boxes.  Table \ref{table:device_errorrates} lists the average error rates for the set of qubits used in these experiments.  These error rates were measured one day before the quantum volume experiments were performed.  Fluctuations in these numbers can occur during the time scale of these experiments, but they are representative of the single-qubit, two-qubit, and measurement errors for each device.  The data from Table \ref{table:device_errorrates} was also used in the noisy simulations of the quantum volume circuits in Table \ref{table:compiled_circuits}.

\begin{table}[h!]
	\centering
	\begin{tabular}{ccccc}
		\toprule
		& Tenerife & Melbourne & Tokyo & Johannesburg \\
		\colrule
		\# Qubits & 5	&	16	&	20 & 20	\\
		$\epsilon_{1Q}$ &	$1.7\times 10^{-3}$	&	$1.6\times 10^{-3}$	&	$1.6\times 10^{-3}$&	$0.4\times 10^{-3}$\\
		$\epsilon_{CX}$	&	$4.7\times 10^{-2}$	&	$3.4\times 10^{-2}$	& 	$2.1\times 10^{-2}$& $1.1\times 10^{-2}$\\
		$\epsilon_{M}$	&	$5.8\times 10^{-2}$	&	$8.7\times 10^{-2}$	&	$3.0\times 10^{-2}$&  $3.9\times 10^{-2}$\\
		\botrule
	\end{tabular}
	\caption{Average error rates for the experimental devices: $\epsilon_{1Q}$ for single-qubit error rates, $\epsilon_{CX}$ for two-qubit error rates, and $\epsilon_{M}$ for measurement.  The averages are taken over the set of qubits from each device that were used in the quantum volume experiments.}
	\label{table:device_errorrates}
\end{table}

\begin{thebibliography}{49}%
\makeatletter
\providecommand \@ifxundefined [1]{%
 \@ifx{#1\undefined}
}%
\providecommand \@ifnum [1]{%
 \ifnum #1\expandafter \@firstoftwo
 \else \expandafter \@secondoftwo
 \fi
}%
\providecommand \@ifx [1]{%
 \ifx #1\expandafter \@firstoftwo
 \else \expandafter \@secondoftwo
 \fi
}%
\providecommand \natexlab [1]{#1}%
\providecommand \enquote  [1]{``#1''}%
\providecommand \bibnamefont  [1]{#1}%
\providecommand \bibfnamefont [1]{#1}%
\providecommand \citenamefont [1]{#1}%
\providecommand \href@noop [0]{\@secondoftwo}%
\providecommand \href [0]{\begingroup \@sanitize@url \@href}%
\providecommand \@href[1]{\@@startlink{#1}\@@href}%
\providecommand \@@href[1]{\endgroup#1\@@endlink}%
\providecommand \@sanitize@url [0]{\catcode `\\12\catcode `\$12\catcode
  `\&12\catcode `\#12\catcode `\^12\catcode `\_12\catcode `\%12\relax}%
\providecommand \@@startlink[1]{}%
\providecommand \@@endlink[0]{}%
\providecommand \url  [0]{\begingroup\@sanitize@url \@url }%
\providecommand \@url [1]{\endgroup\@href {#1}{\urlprefix }}%
\providecommand \urlprefix  [0]{URL }%
\providecommand \Eprint [0]{\href }%
\providecommand \doibase [0]{http://dx.doi.org/}%
\providecommand \selectlanguage [0]{\@gobble}%
\providecommand \bibinfo  [0]{\@secondoftwo}%
\providecommand \bibfield  [0]{\@secondoftwo}%
\providecommand \translation [1]{[#1]}%
\providecommand \BibitemOpen [0]{}%
\providecommand \bibitemStop [0]{}%
\providecommand \bibitemNoStop [0]{.\EOS\space}%
\providecommand \EOS [0]{\spacefactor3000\relax}%
\providecommand \BibitemShut  [1]{\csname bibitem#1\endcsname}%
\let\auto@bib@innerbib\@empty
%</preamble>
\bibitem [{QX()}]{QX}%
  \BibitemOpen
  \href@noop {} {\enquote {\bibinfo {title} {{IBM} {Q} {E}xperience},}\
  }\bibinfo {howpublished}
  {\url{https://quantumexperience.ng.bluemix.net/qx/experience}},\ \bibinfo
  {note} {{L}ast {A}ccessed: 2018-11}\BibitemShut {NoStop}%
\bibitem [{\citenamefont {Friis}\ \emph {et~al.}(2018)\citenamefont {Friis},
  \citenamefont {Marty}, \citenamefont {Maier}, \citenamefont {Hempel},
  \citenamefont {Holz{\"a}pfel}, \citenamefont {Jurcevic}, \citenamefont
  {Plenio}, \citenamefont {Huber}, \citenamefont {Roos}, \citenamefont
  {Blatt},\ and\ \citenamefont {Lanyon}}]{Friis18}%
  \BibitemOpen
  \bibfield  {author} {\bibinfo {author} {\bibfnamefont {N.}~\bibnamefont
  {Friis}}, \bibinfo {author} {\bibfnamefont {O.}~\bibnamefont {Marty}},
  \bibinfo {author} {\bibfnamefont {C.}~\bibnamefont {Maier}}, \bibinfo
  {author} {\bibfnamefont {C.}~\bibnamefont {Hempel}}, \bibinfo {author}
  {\bibfnamefont {M.}~\bibnamefont {Holz{\"a}pfel}}, \bibinfo {author}
  {\bibfnamefont {P.}~\bibnamefont {Jurcevic}}, \bibinfo {author}
  {\bibfnamefont {M.}~\bibnamefont {Plenio}}, \bibinfo {author} {\bibfnamefont
  {M.}~\bibnamefont {Huber}}, \bibinfo {author} {\bibfnamefont
  {C.}~\bibnamefont {Roos}}, \bibinfo {author} {\bibfnamefont {R.}~\bibnamefont
  {Blatt}}, \ and\ \bibinfo {author} {\bibfnamefont {B.}~\bibnamefont
  {Lanyon}},\ }\href {\doibase 10.1103/PhysRevX.8.021012} {\bibfield  {journal}
  {\bibinfo  {journal} {Phys. Rev. X}\ }\textbf {\bibinfo {volume} {8}},\
  \bibinfo {pages} {021012} (\bibinfo {year} {2018})}\BibitemShut {NoStop}%
\bibitem [{\citenamefont {Song}\ \emph {et~al.}(2017)\citenamefont {Song},
  \citenamefont {Xu}, \citenamefont {Liu}, \citenamefont {ping Yang},
  \citenamefont {Zheng}, \citenamefont {Deng}, \citenamefont {Xie},
  \citenamefont {Huang}, \citenamefont {Guo}, \citenamefont {Zhang},
  \citenamefont {Zhang}, \citenamefont {Xu}, \citenamefont {Zheng},
  \citenamefont {Zhu}, \citenamefont {Wang}, \citenamefont {Chen},
  \citenamefont {Lu}, \citenamefont {Han},\ and\ \citenamefont {Pan}}]{Song17}%
  \BibitemOpen
  \bibfield  {author} {\bibinfo {author} {\bibfnamefont {C.}~\bibnamefont
  {Song}}, \bibinfo {author} {\bibfnamefont {K.}~\bibnamefont {Xu}}, \bibinfo
  {author} {\bibfnamefont {W.}~\bibnamefont {Liu}}, \bibinfo {author}
  {\bibfnamefont {C.}~\bibnamefont {ping Yang}}, \bibinfo {author}
  {\bibfnamefont {S.-B.}\ \bibnamefont {Zheng}}, \bibinfo {author}
  {\bibfnamefont {H.}~\bibnamefont {Deng}}, \bibinfo {author} {\bibfnamefont
  {Q.}~\bibnamefont {Xie}}, \bibinfo {author} {\bibfnamefont {K.}~\bibnamefont
  {Huang}}, \bibinfo {author} {\bibfnamefont {Q.}~\bibnamefont {Guo}}, \bibinfo
  {author} {\bibfnamefont {L.}~\bibnamefont {Zhang}}, \bibinfo {author}
  {\bibfnamefont {P.}~\bibnamefont {Zhang}}, \bibinfo {author} {\bibfnamefont
  {D.}~\bibnamefont {Xu}}, \bibinfo {author} {\bibfnamefont {D.}~\bibnamefont
  {Zheng}}, \bibinfo {author} {\bibfnamefont {X.}~\bibnamefont {Zhu}}, \bibinfo
  {author} {\bibfnamefont {H.}~\bibnamefont {Wang}}, \bibinfo {author}
  {\bibfnamefont {Y.-A.}\ \bibnamefont {Chen}}, \bibinfo {author}
  {\bibfnamefont {C.-Y.}\ \bibnamefont {Lu}}, \bibinfo {author} {\bibfnamefont
  {S.}~\bibnamefont {Han}}, \ and\ \bibinfo {author} {\bibfnamefont {J.-W.}\
  \bibnamefont {Pan}},\ }\href {\doibase 10.1103/PhysRevLett.119.180511}
  {\bibfield  {journal} {\bibinfo  {journal} {Phys. Rev. Lett.}\ }\textbf
  {\bibinfo {volume} {119}},\ \bibinfo {pages} {180511} (\bibinfo {year}
  {2017})}\BibitemShut {NoStop}%
\bibitem [{\citenamefont {Preskill}(2018)}]{preskill-nisq}%
  \BibitemOpen
  \bibfield  {author} {\bibinfo {author} {\bibfnamefont {J.}~\bibnamefont
  {Preskill}},\ }\href {https://quantum-journal.org/papers/q-2018-08-06-79/}
  {\bibfield  {journal} {\bibinfo  {journal} {Quantum}\ }\textbf {\bibinfo
  {volume} {2}} (\bibinfo {year} {2018})}\BibitemShut {NoStop}%
\bibitem [{\citenamefont {Magesan}\ \emph {et~al.}(2012)\citenamefont
  {Magesan}, \citenamefont {Gambetta},\ and\ \citenamefont
  {Emerson}}]{magesan_characterizing_2012}%
  \BibitemOpen
  \bibfield  {author} {\bibinfo {author} {\bibfnamefont {E.}~\bibnamefont
  {Magesan}}, \bibinfo {author} {\bibfnamefont {J.~M.}\ \bibnamefont
  {Gambetta}}, \ and\ \bibinfo {author} {\bibfnamefont {J.}~\bibnamefont
  {Emerson}},\ }\href {\doibase 10/tfz} {\bibfield  {journal} {\bibinfo
  {journal} {Physical Review A}\ }\textbf {\bibinfo {volume} {85}},\ \bibinfo
  {pages} {042311} (\bibinfo {year} {2012})}\BibitemShut {NoStop}%
\bibitem [{\citenamefont {Paris}\ and\ \citenamefont
  {\v{R}eh\'a\v{c}ek}(2004)}]{paris_quantum_2004}%
  \BibitemOpen
  \bibinfo {editor} {\bibfnamefont {M.~G.~A.}\ \bibnamefont {Paris}}\ and\
  \bibinfo {editor} {\bibfnamefont {J.}~\bibnamefont {\v{R}eh\'a\v{c}ek}},\
  eds.,\ \href {http://www.springer.com/us/book/9783540223290} {\emph {\bibinfo
  {title} {Quantum {State} {Estimation}}}},\ Lecture {Notes} in {Physics}\
  (\bibinfo  {publisher} {Springer-Verlag},\ \bibinfo {address} {Berlin
  Heidelberg},\ \bibinfo {year} {2004})\BibitemShut {NoStop}%
\bibitem [{\citenamefont {Merkel}\ \emph {et~al.}(2013)\citenamefont {Merkel},
  \citenamefont {Gambetta}, \citenamefont {Smolin}, \citenamefont {Poletto},
  \citenamefont {C\'orcoles}, \citenamefont {Johnson}, \citenamefont {Ryan},\
  and\ \citenamefont {Steffen}}]{merkel_self-consistent_2013}%
  \BibitemOpen
  \bibfield  {author} {\bibinfo {author} {\bibfnamefont {S.~T.}\ \bibnamefont
  {Merkel}}, \bibinfo {author} {\bibfnamefont {J.~M.}\ \bibnamefont
  {Gambetta}}, \bibinfo {author} {\bibfnamefont {J.~A.}\ \bibnamefont
  {Smolin}}, \bibinfo {author} {\bibfnamefont {S.}~\bibnamefont {Poletto}},
  \bibinfo {author} {\bibfnamefont {A.~D.}\ \bibnamefont {C\'orcoles}},
  \bibinfo {author} {\bibfnamefont {B.~R.}\ \bibnamefont {Johnson}}, \bibinfo
  {author} {\bibfnamefont {C.~A.}\ \bibnamefont {Ryan}}, \ and\ \bibinfo
  {author} {\bibfnamefont {M.}~\bibnamefont {Steffen}},\ }\href {\doibase
  10/gdcctv} {\bibfield  {journal} {\bibinfo  {journal} {Physical Review A}\
  }\textbf {\bibinfo {volume} {87}},\ \bibinfo {pages} {062119} (\bibinfo
  {year} {2013})}\BibitemShut {NoStop}%
\bibitem [{\citenamefont {Blume-Kohout}\ \emph {et~al.}(2017)\citenamefont
  {Blume-Kohout}, \citenamefont {Gamble}, \citenamefont {Nielsen},
  \citenamefont {Rudinger}, \citenamefont {Mizrahi}, \citenamefont {Fortier},\
  and\ \citenamefont {Maunz}}]{blume-kohout_demonstration_2017}%
  \BibitemOpen
  \bibfield  {author} {\bibinfo {author} {\bibfnamefont {R.}~\bibnamefont
  {Blume-Kohout}}, \bibinfo {author} {\bibfnamefont {J.~K.}\ \bibnamefont
  {Gamble}}, \bibinfo {author} {\bibfnamefont {E.}~\bibnamefont {Nielsen}},
  \bibinfo {author} {\bibfnamefont {K.}~\bibnamefont {Rudinger}}, \bibinfo
  {author} {\bibfnamefont {J.}~\bibnamefont {Mizrahi}}, \bibinfo {author}
  {\bibfnamefont {K.}~\bibnamefont {Fortier}}, \ and\ \bibinfo {author}
  {\bibfnamefont {P.}~\bibnamefont {Maunz}},\ }\href {\doibase 10/f9qz5p}
  {\bibfield  {journal} {\bibinfo  {journal} {Nature Communications}\ }\textbf
  {\bibinfo {volume} {8}},\ \bibinfo {pages} {14485} (\bibinfo {year}
  {2017})}\BibitemShut {NoStop}%
\bibitem [{\citenamefont {McKay}\ \emph {et~al.}(2017)\citenamefont {McKay},
  \citenamefont {Sheldon}, \citenamefont {Smolin}, \citenamefont {Chow},\ and\
  \citenamefont {Gambetta}}]{McKay17}%
  \BibitemOpen
  \bibfield  {author} {\bibinfo {author} {\bibfnamefont {D.~C.}\ \bibnamefont
  {McKay}}, \bibinfo {author} {\bibfnamefont {S.}~\bibnamefont {Sheldon}},
  \bibinfo {author} {\bibfnamefont {J.~A.}\ \bibnamefont {Smolin}}, \bibinfo
  {author} {\bibfnamefont {J.~M.}\ \bibnamefont {Chow}}, \ and\ \bibinfo
  {author} {\bibfnamefont {J.~M.}\ \bibnamefont {Gambetta}},\ }\href
  {https://arxiv.org/abs/1712.06550} {\bibfield  {journal} {\bibinfo  {journal}
  {arXiv preprint}\ }\textbf {\bibinfo {volume} {arXiv:1712.06550}} (\bibinfo
  {year} {2017})}\BibitemShut {NoStop}%
\bibitem [{\citenamefont {Takita}\ \emph {et~al.}(2017)\citenamefont {Takita},
  \citenamefont {Cross}, \citenamefont {C\'orcoles}, \citenamefont {Chow},\
  and\ \citenamefont {Gambetta}}]{Takita17}%
  \BibitemOpen
  \bibfield  {author} {\bibinfo {author} {\bibfnamefont {M.}~\bibnamefont
  {Takita}}, \bibinfo {author} {\bibfnamefont {A.~W.}\ \bibnamefont {Cross}},
  \bibinfo {author} {\bibfnamefont {A.~D.}\ \bibnamefont {C\'orcoles}},
  \bibinfo {author} {\bibfnamefont {J.~M.}\ \bibnamefont {Chow}}, \ and\
  \bibinfo {author} {\bibfnamefont {J.~M.}\ \bibnamefont {Gambetta}},\ }\href
  {\doibase 10.1103/PhysRevLett.119.180501} {\bibfield  {journal} {\bibinfo
  {journal} {Phys. Rev. Lett.}\ }\textbf {\bibinfo {volume} {119}},\ \bibinfo
  {pages} {180501} (\bibinfo {year} {2017})}\BibitemShut {NoStop}%
\bibitem [{\citenamefont {Dongarra}\ \emph {et~al.}(2003)\citenamefont
  {Dongarra}, \citenamefont {Luszczek},\ and\ \citenamefont
  {Petitet}}]{LINPACK}%
  \BibitemOpen
  \bibfield  {author} {\bibinfo {author} {\bibfnamefont {J.~J.}\ \bibnamefont
  {Dongarra}}, \bibinfo {author} {\bibfnamefont {P.}~\bibnamefont {Luszczek}},
  \ and\ \bibinfo {author} {\bibfnamefont {A.}~\bibnamefont {Petitet}},\ }\href
  {\doibase 10/c242wc} {\bibfield  {journal} {\bibinfo  {journal} {Concurrency
  and Computation: Practice and Experience}\ }\textbf {\bibinfo {volume}
  {15}},\ \bibinfo {pages} {803} (\bibinfo {year} {2003})}\BibitemShut
  {NoStop}%
\bibitem [{\citenamefont {Dongarra}\ \emph {et~al.}(2015)\citenamefont
  {Dongarra}, \citenamefont {Heroux},\ and\ \citenamefont {Luszczek}}]{HPCG}%
  \BibitemOpen
  \bibfield  {author} {\bibinfo {author} {\bibfnamefont {J.}~\bibnamefont
  {Dongarra}}, \bibinfo {author} {\bibfnamefont {M.~A.}\ \bibnamefont
  {Heroux}}, \ and\ \bibinfo {author} {\bibfnamefont {P.}~\bibnamefont
  {Luszczek}},\ }\href@noop {} {\emph {\bibinfo {title} {{HPCG} {Benchmark}: a
  {New} {Metric} for {Ranking} {High} {Performance} {Computing} {Systems}}}},\
  \bibinfo {type} {Technical Report}\ \bibinfo {number} {UT-EECS-15-736}\
  (\bibinfo  {institution} {Electrical Engineering and Computer Science
  Department, Knoxville, Tennessee},\ \bibinfo {year} {2015})\BibitemShut
  {NoStop}%
\bibitem [{\citenamefont {Adams}(2014)}]{HPGMG}%
  \BibitemOpen
  \bibfield  {author} {\bibinfo {author} {\bibfnamefont {M.}~\bibnamefont
  {Adams}},\ }\href {https://escholarship.org/uc/item/00r9w79m} {\emph
  {\bibinfo {title} {{HPGMG} 1.0: {A} {Benchmark} for {Ranking} {High}
  {Performance} {Computing} {Systems}}}},\ \bibinfo {type} {Technical Report}\
  \bibinfo {number} {LBNL 6630E}\ (\bibinfo  {institution} {LBNL},\ \bibinfo
  {year} {2014})\BibitemShut {NoStop}%
\bibitem [{\citenamefont {Bishop}\ \emph {et~al.}(2017)\citenamefont {Bishop},
  \citenamefont {Bravyi}, \citenamefont {Cross}, \citenamefont {Gambetta},\
  and\ \citenamefont {Smolin}}]{qvolume}%
  \BibitemOpen
  \bibfield  {author} {\bibinfo {author} {\bibfnamefont {L.~S.}\ \bibnamefont
  {Bishop}}, \bibinfo {author} {\bibfnamefont {S.}~\bibnamefont {Bravyi}},
  \bibinfo {author} {\bibfnamefont {A.}~\bibnamefont {Cross}}, \bibinfo
  {author} {\bibfnamefont {J.~M.}\ \bibnamefont {Gambetta}}, \ and\ \bibinfo
  {author} {\bibfnamefont {J.~A.}\ \bibnamefont {Smolin}},\ }\href
  {https://dal.objectstorage.open.softlayer.com/v1/AUTH_039c3bf6e6e54d76b8e66152e2f87877/community-documents/quatnum-volumehp08co1vbo0cc8fr.pdf}
  {\enquote {\bibinfo {title} {Quantum volume},}\ } (\bibinfo {year}
  {2017})\BibitemShut {NoStop}%
\bibitem [{\citenamefont {Moll}\ \emph {et~al.}(2018)\citenamefont {Moll},
  \citenamefont {Barkoutsos}, \citenamefont {Bishop}, \citenamefont {Chow},
  \citenamefont {Cross}, \citenamefont {Egger}, \citenamefont {Filipp},
  \citenamefont {Fuhrer}, \citenamefont {Gambetta}, \citenamefont {Ganzhorn},
  \citenamefont {Kandala}, \citenamefont {Mezzacapo}, \citenamefont {Muller},
  \citenamefont {Riess}, \citenamefont {Salis}, \citenamefont {Smolin},
  \citenamefont {Tavernelli},\ and\ \citenamefont {Temme}}]{Moll18}%
  \BibitemOpen
  \bibfield  {author} {\bibinfo {author} {\bibfnamefont {N.}~\bibnamefont
  {Moll}}, \bibinfo {author} {\bibfnamefont {P.}~\bibnamefont {Barkoutsos}},
  \bibinfo {author} {\bibfnamefont {L.~S.}\ \bibnamefont {Bishop}}, \bibinfo
  {author} {\bibfnamefont {J.~M.}\ \bibnamefont {Chow}}, \bibinfo {author}
  {\bibfnamefont {A.}~\bibnamefont {Cross}}, \bibinfo {author} {\bibfnamefont
  {D.~J.}\ \bibnamefont {Egger}}, \bibinfo {author} {\bibfnamefont
  {S.}~\bibnamefont {Filipp}}, \bibinfo {author} {\bibfnamefont
  {A.}~\bibnamefont {Fuhrer}}, \bibinfo {author} {\bibfnamefont {J.~M.}\
  \bibnamefont {Gambetta}}, \bibinfo {author} {\bibfnamefont {M.}~\bibnamefont
  {Ganzhorn}}, \bibinfo {author} {\bibfnamefont {A.}~\bibnamefont {Kandala}},
  \bibinfo {author} {\bibfnamefont {A.}~\bibnamefont {Mezzacapo}}, \bibinfo
  {author} {\bibfnamefont {P.}~\bibnamefont {Muller}}, \bibinfo {author}
  {\bibfnamefont {W.}~\bibnamefont {Riess}}, \bibinfo {author} {\bibfnamefont
  {G.}~\bibnamefont {Salis}}, \bibinfo {author} {\bibfnamefont
  {J.}~\bibnamefont {Smolin}}, \bibinfo {author} {\bibfnamefont
  {I.}~\bibnamefont {Tavernelli}}, \ and\ \bibinfo {author} {\bibfnamefont
  {K.}~\bibnamefont {Temme}},\ }\href {\doibase 10.1088/2058-9565/aab822}
  {\bibfield  {journal} {\bibinfo  {journal} {Quantum Sci. Technol.}\ }\textbf
  {\bibinfo {volume} {3}},\ \bibinfo {pages} {030503} (\bibinfo {year}
  {2018})}\BibitemShut {NoStop}%
\bibitem [{\citenamefont {Nielsen}\ and\ \citenamefont {Chuang}(2000)}]{NC}%
  \BibitemOpen
  \bibfield  {author} {\bibinfo {author} {\bibfnamefont {M.}~\bibnamefont
  {Nielsen}}\ and\ \bibinfo {author} {\bibfnamefont {I.}~\bibnamefont
  {Chuang}},\ }\href@noop {} {\emph {\bibinfo {title} {Quantum computation and
  quantum information}}}\ (\bibinfo  {publisher} {Cambridge},\ \bibinfo {year}
  {2000})\BibitemShut {NoStop}%
\bibitem [{\citenamefont {Boixo}\ \emph {et~al.}(2018)\citenamefont {Boixo},
  \citenamefont {Isakov}, \citenamefont {Smelyanskiy}, \citenamefont {Babbush},
  \citenamefont {Ding}, \citenamefont {Jiang}, \citenamefont {Bremner},
  \citenamefont {Martinis},\ and\ \citenamefont {Neven}}]{B18}%
  \BibitemOpen
  \bibfield  {author} {\bibinfo {author} {\bibfnamefont {S.}~\bibnamefont
  {Boixo}}, \bibinfo {author} {\bibfnamefont {S.~V.}\ \bibnamefont {Isakov}},
  \bibinfo {author} {\bibfnamefont {V.~N.}\ \bibnamefont {Smelyanskiy}},
  \bibinfo {author} {\bibfnamefont {R.}~\bibnamefont {Babbush}}, \bibinfo
  {author} {\bibfnamefont {N.}~\bibnamefont {Ding}}, \bibinfo {author}
  {\bibfnamefont {Z.}~\bibnamefont {Jiang}}, \bibinfo {author} {\bibfnamefont
  {M.~J.}\ \bibnamefont {Bremner}}, \bibinfo {author} {\bibfnamefont {J.~M.}\
  \bibnamefont {Martinis}}, \ and\ \bibinfo {author} {\bibfnamefont
  {H.}~\bibnamefont {Neven}},\ }\href {\doibase 10.1038/s41567-018-0124-x}
  {\bibfield  {journal} {\bibinfo  {journal} {Nat. Phys.}\ }\textbf {\bibinfo
  {volume} {14}},\ \bibinfo {pages} {595} (\bibinfo {year} {2018})}\BibitemShut
  {NoStop}%
\bibitem [{\citenamefont {Farhi}\ \emph {et~al.}(2014)\citenamefont {Farhi},
  \citenamefont {Goldstone},\ and\ \citenamefont {Gutmann}}]{FGG14}%
  \BibitemOpen
  \bibfield  {author} {\bibinfo {author} {\bibfnamefont {E.}~\bibnamefont
  {Farhi}}, \bibinfo {author} {\bibfnamefont {J.}~\bibnamefont {Goldstone}}, \
  and\ \bibinfo {author} {\bibfnamefont {S.}~\bibnamefont {Gutmann}},\ }\href
  {https://arxiv.org/abs/1411.4028} {\bibfield  {journal} {\bibinfo  {journal}
  {arXiv preprint}\ }\textbf {\bibinfo {volume} {arXiv:1411.4028}} (\bibinfo
  {year} {2014})}\BibitemShut {NoStop}%
\bibitem [{\citenamefont {McClean}\ \emph {et~al.}(2016)\citenamefont
  {McClean}, \citenamefont {Romero}, \citenamefont {Babbush},\ and\
  \citenamefont {Aspuru-Guzik}}]{McClean16}%
  \BibitemOpen
  \bibfield  {author} {\bibinfo {author} {\bibfnamefont {J.~R.}\ \bibnamefont
  {McClean}}, \bibinfo {author} {\bibfnamefont {J.}~\bibnamefont {Romero}},
  \bibinfo {author} {\bibfnamefont {R.}~\bibnamefont {Babbush}}, \ and\
  \bibinfo {author} {\bibfnamefont {A.}~\bibnamefont {Aspuru-Guzik}},\ }\href
  {\doibase 10.1088/1367-2630/18/2/023023} {\bibfield  {journal} {\bibinfo
  {journal} {New J. Phys.}\ }\textbf {\bibinfo {volume} {18}},\ \bibinfo
  {pages} {023023} (\bibinfo {year} {2016})}\BibitemShut {NoStop}%
\bibitem [{\citenamefont {Yung}\ \emph {et~al.}(2014)\citenamefont {Yung},
  \citenamefont {Casanova}, \citenamefont {Mezzacapo}, \citenamefont {McClean},
  \citenamefont {Lamata}, \citenamefont {Aspuru-Guzik},\ and\ \citenamefont
  {Solano}}]{Yung14}%
  \BibitemOpen
  \bibfield  {author} {\bibinfo {author} {\bibfnamefont {M.-H.}\ \bibnamefont
  {Yung}}, \bibinfo {author} {\bibfnamefont {J.}~\bibnamefont {Casanova}},
  \bibinfo {author} {\bibfnamefont {A.}~\bibnamefont {Mezzacapo}}, \bibinfo
  {author} {\bibfnamefont {J.}~\bibnamefont {McClean}}, \bibinfo {author}
  {\bibfnamefont {L.}~\bibnamefont {Lamata}}, \bibinfo {author} {\bibfnamefont
  {A.}~\bibnamefont {Aspuru-Guzik}}, \ and\ \bibinfo {author} {\bibfnamefont
  {E.}~\bibnamefont {Solano}},\ }\href
  {https://www.nature.com/articles/srep03589} {\bibfield  {journal} {\bibinfo
  {journal} {Sci. Rep.}\ }\textbf {\bibinfo {volume} {4}} (\bibinfo {year}
  {2014})}\BibitemShut {NoStop}%
\bibitem [{\citenamefont {Aaronson}\ and\ \citenamefont {Chen}(2016)}]{AC16}%
  \BibitemOpen
  \bibfield  {author} {\bibinfo {author} {\bibfnamefont {S.}~\bibnamefont
  {Aaronson}}\ and\ \bibinfo {author} {\bibfnamefont {L.}~\bibnamefont
  {Chen}},\ }\href {https://arxiv.org/abs/1612.05903} {\bibfield  {journal}
  {\bibinfo  {journal} {arXiv preprint}\ }\textbf {\bibinfo {volume}
  {arXiv:1612.05903}} (\bibinfo {year} {2016})}\BibitemShut {NoStop}%
\bibitem [{\citenamefont {Horodecki}\ \emph {et~al.}(1999)\citenamefont
  {Horodecki}, \citenamefont {Horodecki},\ and\ \citenamefont
  {Horodecki}}]{horodecki_general_1999}%
  \BibitemOpen
  \bibfield  {author} {\bibinfo {author} {\bibfnamefont {M.}~\bibnamefont
  {Horodecki}}, \bibinfo {author} {\bibfnamefont {P.}~\bibnamefont
  {Horodecki}}, \ and\ \bibinfo {author} {\bibfnamefont {R.}~\bibnamefont
  {Horodecki}},\ }\href {\doibase 10/b9tmqn} {\bibfield  {journal} {\bibinfo
  {journal} {Physical Review A}\ }\textbf {\bibinfo {volume} {60}},\ \bibinfo
  {pages} {1888} (\bibinfo {year} {1999})}\BibitemShut {NoStop}%
\bibitem [{\citenamefont {Chen}\ \emph
  {et~al.}(2018{\natexlab{a}})\citenamefont {Chen}, \citenamefont {Zhou},
  \citenamefont {Xue}, \citenamefont {Yang}, \citenamefont {Guo},\ and\
  \citenamefont {Guo}}]{chen_64-qubit_2018}%
  \BibitemOpen
  \bibfield  {author} {\bibinfo {author} {\bibfnamefont {Z.-Y.}\ \bibnamefont
  {Chen}}, \bibinfo {author} {\bibfnamefont {Q.}~\bibnamefont {Zhou}}, \bibinfo
  {author} {\bibfnamefont {C.}~\bibnamefont {Xue}}, \bibinfo {author}
  {\bibfnamefont {X.}~\bibnamefont {Yang}}, \bibinfo {author} {\bibfnamefont
  {G.-C.}\ \bibnamefont {Guo}}, \ and\ \bibinfo {author} {\bibfnamefont
  {G.-P.}\ \bibnamefont {Guo}},\ }\href {\doibase 10/gffc8t} {\bibfield
  {journal} {\bibinfo  {journal} {Science Bulletin}\ }\textbf {\bibinfo
  {volume} {63}},\ \bibinfo {pages} {964} (\bibinfo {year}
  {2018}{\natexlab{a}})},\ \bibinfo {note} {arXiv: 1802.06952}\BibitemShut
  {NoStop}%
\bibitem [{\citenamefont {Chen}\ \emph
  {et~al.}(2018{\natexlab{b}})\citenamefont {Chen}, \citenamefont {Zhang},
  \citenamefont {Huang}, \citenamefont {Newman},\ and\ \citenamefont
  {Shi}}]{chen_classical_2018}%
  \BibitemOpen
  \bibfield  {author} {\bibinfo {author} {\bibfnamefont {J.}~\bibnamefont
  {Chen}}, \bibinfo {author} {\bibfnamefont {F.}~\bibnamefont {Zhang}},
  \bibinfo {author} {\bibfnamefont {C.}~\bibnamefont {Huang}}, \bibinfo
  {author} {\bibfnamefont {M.}~\bibnamefont {Newman}}, \ and\ \bibinfo {author}
  {\bibfnamefont {Y.}~\bibnamefont {Shi}},\ }\href
  {http://arxiv.org/abs/1805.01450} {\bibfield  {journal} {\bibinfo  {journal}
  {arXiv:1805.01450 [quant-ph]}\ } (\bibinfo {year} {2018}{\natexlab{b}})},\
  \bibinfo {note} {arXiv: 1805.01450}\BibitemShut {NoStop}%
\bibitem [{\citenamefont {Li}\ \emph {et~al.}(2018)\citenamefont {Li},
  \citenamefont {Wu}, \citenamefont {Ying}, \citenamefont {Sun},\ and\
  \citenamefont {Yang}}]{li_quantum_2018}%
  \BibitemOpen
  \bibfield  {author} {\bibinfo {author} {\bibfnamefont {R.}~\bibnamefont
  {Li}}, \bibinfo {author} {\bibfnamefont {B.}~\bibnamefont {Wu}}, \bibinfo
  {author} {\bibfnamefont {M.}~\bibnamefont {Ying}}, \bibinfo {author}
  {\bibfnamefont {X.}~\bibnamefont {Sun}}, \ and\ \bibinfo {author}
  {\bibfnamefont {G.}~\bibnamefont {Yang}},\ }\href
  {http://arxiv.org/abs/1804.04797} {\bibfield  {journal} {\bibinfo  {journal}
  {arXiv:1804.04797 [quant-ph]}\ } (\bibinfo {year} {2018})},\ \bibinfo {note}
  {arXiv: 1804.04797}\BibitemShut {NoStop}%
\bibitem [{\citenamefont {Pednault}\ \emph {et~al.}(2017)\citenamefont
  {Pednault}, \citenamefont {Gunnels}, \citenamefont {Nannicini}, \citenamefont
  {Horesh}, \citenamefont {Magerlein}, \citenamefont {Solomonik},\ and\
  \citenamefont {Wisnieff}}]{pednault_breaking_2017}%
  \BibitemOpen
  \bibfield  {author} {\bibinfo {author} {\bibfnamefont {E.}~\bibnamefont
  {Pednault}}, \bibinfo {author} {\bibfnamefont {J.~A.}\ \bibnamefont
  {Gunnels}}, \bibinfo {author} {\bibfnamefont {G.}~\bibnamefont {Nannicini}},
  \bibinfo {author} {\bibfnamefont {L.}~\bibnamefont {Horesh}}, \bibinfo
  {author} {\bibfnamefont {T.}~\bibnamefont {Magerlein}}, \bibinfo {author}
  {\bibfnamefont {E.}~\bibnamefont {Solomonik}}, \ and\ \bibinfo {author}
  {\bibfnamefont {R.}~\bibnamefont {Wisnieff}},\ }\href
  {http://arxiv.org/abs/1710.05867} {\bibfield  {journal} {\bibinfo  {journal}
  {arXiv:1710.05867 [quant-ph]}\ } (\bibinfo {year} {2017})},\ \bibinfo {note}
  {arXiv: 1710.05867}\BibitemShut {NoStop}%
\bibitem [{\citenamefont {Boixo}\ \emph {et~al.}(2017)\citenamefont {Boixo},
  \citenamefont {Isakov}, \citenamefont {Smelyanskiy},\ and\ \citenamefont
  {Neven}}]{boixo_simulation_2017}%
  \BibitemOpen
  \bibfield  {author} {\bibinfo {author} {\bibfnamefont {S.}~\bibnamefont
  {Boixo}}, \bibinfo {author} {\bibfnamefont {S.~V.}\ \bibnamefont {Isakov}},
  \bibinfo {author} {\bibfnamefont {V.~N.}\ \bibnamefont {Smelyanskiy}}, \ and\
  \bibinfo {author} {\bibfnamefont {H.}~\bibnamefont {Neven}},\ }\href
  {http://arxiv.org/abs/1712.05384} {\bibfield  {journal} {\bibinfo  {journal}
  {arXiv:1712.05384 [quant-ph]}\ } (\bibinfo {year} {2017})},\ \bibinfo {note}
  {arXiv: 1712.05384}\BibitemShut {NoStop}%
\bibitem [{\citenamefont {H{\"a}ner}\ \emph {et~al.}(2016)\citenamefont
  {H{\"a}ner}, \citenamefont {Steiger}, \citenamefont {Smelyanskiy},\ and\
  \citenamefont {Troyer}}]{haner_high_2016}%
  \BibitemOpen
  \bibfield  {author} {\bibinfo {author} {\bibfnamefont {T.}~\bibnamefont
  {H{\"a}ner}}, \bibinfo {author} {\bibfnamefont {D.~S.}\ \bibnamefont
  {Steiger}}, \bibinfo {author} {\bibfnamefont {M.}~\bibnamefont
  {Smelyanskiy}}, \ and\ \bibinfo {author} {\bibfnamefont {M.}~\bibnamefont
  {Troyer}},\ }\href {http://arxiv.org/abs/1604.06460} {\bibfield  {journal}
  {\bibinfo  {journal} {arXiv:1604.06460 [quant-ph]}\ } (\bibinfo {year}
  {2016})},\ \bibinfo {note} {arXiv: 1604.06460}\BibitemShut {NoStop}%
\bibitem [{\citenamefont {Smelyanskiy}\ \emph {et~al.}(2016)\citenamefont
  {Smelyanskiy}, \citenamefont {Sawaya},\ and\ \citenamefont
  {Aspuru-Guzik}}]{smelyanskiy_qhipster:_2016}%
  \BibitemOpen
  \bibfield  {author} {\bibinfo {author} {\bibfnamefont {M.}~\bibnamefont
  {Smelyanskiy}}, \bibinfo {author} {\bibfnamefont {N.~P.~D.}\ \bibnamefont
  {Sawaya}}, \ and\ \bibinfo {author} {\bibfnamefont {A.}~\bibnamefont
  {Aspuru-Guzik}},\ }\href {http://arxiv.org/abs/1601.07195} {\bibfield
  {journal} {\bibinfo  {journal} {arXiv:1601.07195 [quant-ph]}\ } (\bibinfo
  {year} {2016})},\ \bibinfo {note} {arXiv: 1601.07195}\BibitemShut {NoStop}%
\bibitem [{\citenamefont {Markov}\ and\ \citenamefont
  {Shi}(2008)}]{markov_simulating_2008}%
  \BibitemOpen
  \bibfield  {author} {\bibinfo {author} {\bibfnamefont {I.}~\bibnamefont
  {Markov}}\ and\ \bibinfo {author} {\bibfnamefont {Y.}~\bibnamefont {Shi}},\
  }\href {\doibase 10/fs9htp} {\bibfield  {journal} {\bibinfo  {journal} {SIAM
  Journal on Computing}\ }\textbf {\bibinfo {volume} {38}},\ \bibinfo {pages}
  {963} (\bibinfo {year} {2008})}\BibitemShut {NoStop}%
\bibitem [{ten()}]{tenerife}%
  \BibitemOpen
  \href@noop {} {\enquote {\bibinfo {title} {5-qubit backend: {IBM Q} team,
  "{IBM Q} 5 {T}enerife backend specification v1.3.0,” (2018).}}\ }\bibinfo
  {howpublished} {\url{https://ibm.biz/qiskit-tenerife}},\ \bibinfo {note}
  {{L}ast {A}ccessed: 2018-11}\BibitemShut {NoStop}%
\bibitem [{mel()}]{melbourne}%
  \BibitemOpen
  \href@noop {} {\enquote {\bibinfo {title} {16-qubit backend: {IBM Q} team,
  "{IBM Q} 16 {M}elbourne backend specification v1.0.0,” (2018).}}\ }\bibinfo
  {howpublished} {\url{https://ibm.biz/qiskit-melbourne}},\ \bibinfo {note}
  {{L}ast {A}ccessed: 2018-11}\BibitemShut {NoStop}%
\bibitem [{\citenamefont {Bullock}\ and\ \citenamefont {Markov}(2003)}]{BM03}%
  \BibitemOpen
  \bibfield  {author} {\bibinfo {author} {\bibfnamefont {S.}~\bibnamefont
  {Bullock}}\ and\ \bibinfo {author} {\bibfnamefont {I.}~\bibnamefont
  {Markov}},\ }\href {\doibase 10.1103/PhysRevA.68.012318} {\bibfield
  {journal} {\bibinfo  {journal} {Phys. Rev. A}\ }\textbf {\bibinfo {volume}
  {68}},\ \bibinfo {pages} {012318} (\bibinfo {year} {2003})}\BibitemShut
  {NoStop}%
\bibitem [{\citenamefont {Shende}\ \emph {et~al.}(2004)\citenamefont {Shende},
  \citenamefont {Markov},\ and\ \citenamefont {Bullock}}]{SMB04}%
  \BibitemOpen
  \bibfield  {author} {\bibinfo {author} {\bibfnamefont {V.}~\bibnamefont
  {Shende}}, \bibinfo {author} {\bibfnamefont {I.}~\bibnamefont {Markov}}, \
  and\ \bibinfo {author} {\bibfnamefont {S.}~\bibnamefont {Bullock}},\ }\href
  {\doibase 10.1103/PhysRevA.69.062321} {\bibfield  {journal} {\bibinfo
  {journal} {Phys. Rev. A}\ }\textbf {\bibinfo {volume} {69}},\ \bibinfo
  {pages} {062321} (\bibinfo {year} {2004})}\BibitemShut {NoStop}%
\bibitem [{\citenamefont {Abraham}\ \emph {et~al.}(2019)\citenamefont
  {Abraham}, \citenamefont {Akhalwaya}, \citenamefont {Aleksandrowicz},
  \citenamefont {Alexander}, \citenamefont {Alexandrowics}, \citenamefont
  {Arbel}, \citenamefont {Asfaw}, \citenamefont {Azaustre}, \citenamefont
  {Barkoutsos}, \citenamefont {Barron}, \citenamefont {Bello}, \citenamefont
  {Ben-Haim}, \citenamefont {Bishop}, \citenamefont {Bosch}, \citenamefont
  {Bucher}, \citenamefont {CZ}, \citenamefont {Cabrera}, \citenamefont
  {Calpin}, \citenamefont {Capelluto}, \citenamefont {Carballo}, \citenamefont
  {Chen}, \citenamefont {Chen}, \citenamefont {Chen}, \citenamefont {Chow},
  \citenamefont {Claus}, \citenamefont {Cross}, \citenamefont {Cross},
  \citenamefont {Cruz-Benito}, \citenamefont {Culver}, \citenamefont
  {C{\'o}rcoles-Gonzales}, \citenamefont {Dague}, \citenamefont {Dartiailh},
  \citenamefont {Davila}, \citenamefont {Ding}, \citenamefont {Dumitrescu},
  \citenamefont {Dumon}, \citenamefont {Duran}, \citenamefont {Eendebak},
  \citenamefont {Egger}, \citenamefont {Everitt}, \citenamefont
  {Fern{\'a}ndez}, \citenamefont {Frisch}, \citenamefont {Fuhrer},
  \citenamefont {Gacon}, \citenamefont {Gadi}, \citenamefont {Gago},
  \citenamefont {Gambetta}, \citenamefont {Garcia}, \citenamefont {Garion},
  \citenamefont {Gawel-Kus}, \citenamefont {Gil}, \citenamefont
  {Gomez-Mosquera}, \citenamefont {de~la Puente~Gonz{\'a}lez}, \citenamefont
  {Greenberg}, \citenamefont {Gunnels}, \citenamefont {Haide}, \citenamefont
  {Hamamura}, \citenamefont {Havlicek}, \citenamefont {Hellmers}, \citenamefont
  {Herok}, \citenamefont {Horii}, \citenamefont {Howington}, \citenamefont
  {Hu}, \citenamefont {Hu}, \citenamefont {Imai}, \citenamefont {Imamichi},
  \citenamefont {Iten}, \citenamefont {Itoko}, \citenamefont {Javadi-Abhari},
  \citenamefont {Jessica}, \citenamefont {Johns}, \citenamefont {Kanazawa},
  \citenamefont {Karazeev}, \citenamefont {Kassebaum}, \citenamefont
  {Krishnan}, \citenamefont {Krsulich}, \citenamefont {Kus}, \citenamefont
  {LaRose}, \citenamefont {Lambert}, \citenamefont {Latone}, \citenamefont
  {Lawrence}, \citenamefont {Liu}, \citenamefont {Mac}, \citenamefont {Maeng},
  \citenamefont {Malyshev}, \citenamefont {Marecek}, \citenamefont {Marques},
  \citenamefont {Mathews}, \citenamefont {Matsuo}, \citenamefont {McClure},
  \citenamefont {McGarry}, \citenamefont {McKay}, \citenamefont {Meesala},
  \citenamefont {Mezzacapo}, \citenamefont {Midha}, \citenamefont {Minev},
  \citenamefont {Murali}, \citenamefont {M{\"u}ggenburg}, \citenamefont
  {Nadlinger}, \citenamefont {Nannicini}, \citenamefont {Nation}, \citenamefont
  {Naveh}, \citenamefont {Nick-Singstock}, \citenamefont {Niroula},
  \citenamefont {Norlen}, \citenamefont {O'Riordan}, \citenamefont {Oud},
  \citenamefont {Padilha}, \citenamefont {Paik}, \citenamefont {Perriello},
  \citenamefont {Phan}, \citenamefont {Pistoia}, \citenamefont
  {Pozas-iKerstjens}, \citenamefont {Prutyanov}, \citenamefont {P{\'e}rez},
  \citenamefont {Quintiii}, \citenamefont {Raymond}, \citenamefont {Redondo},
  \citenamefont {Reuter}, \citenamefont {Rodr{\'\i}guez}, \citenamefont {Ryu},
  \citenamefont {Sandberg}, \citenamefont {Sathaye}, \citenamefont {Schmitt},
  \citenamefont {Schnabel}, \citenamefont {Scholten}, \citenamefont {Schoute},
  \citenamefont {Sertage}, \citenamefont {Shi}, \citenamefont {Silva},
  \citenamefont {Siraichi}, \citenamefont {Sivarajah}, \citenamefont {Smolin},
  \citenamefont {Soeken}, \citenamefont {Steenken}, \citenamefont
  {Stypulkoski}, \citenamefont {Takahashi}, \citenamefont {Taylor},
  \citenamefont {Taylour}, \citenamefont {Thomas}, \citenamefont {Tillet},
  \citenamefont {Tod}, \citenamefont {de~la Torre}, \citenamefont {Trabing},
  \citenamefont {Treinish}, \citenamefont {TrishaPe}, \citenamefont {Turner},
  \citenamefont {Vaknin}, \citenamefont {Valcarce}, \citenamefont {Varchon},
  \citenamefont {Vogt-Lee}, \citenamefont {Vuillot}, \citenamefont {Weaver},
  \citenamefont {Wieczorek}, \citenamefont {Wildstrom}, \citenamefont {Wille},
  \citenamefont {Winston}, \citenamefont {Woehr}, \citenamefont {Woerner},
  \citenamefont {Woo}, \citenamefont {Wood}, \citenamefont {Wood},
  \citenamefont {Wood}, \citenamefont {Wootton}, \citenamefont {Yeralin},
  \citenamefont {Yu}, \citenamefont {Zdanski}, \citenamefont {Zoufalc},
  \citenamefont {azulehner}, \citenamefont {drholmie}, \citenamefont
  {fanizzamarco}, \citenamefont {kanejess}, \citenamefont {klinvill},
  \citenamefont {merav aharoni}, \citenamefont {ordmoj}, \citenamefont
  {tigerjack}, \citenamefont {yang.luh},\ and\ \citenamefont
  {yotamvakninibm}}]{Qiskit}%
  \BibitemOpen
  \bibfield  {author} {\bibinfo {author} {\bibfnamefont {H.}~\bibnamefont
  {Abraham}}, \bibinfo {author} {\bibfnamefont {I.~Y.}\ \bibnamefont
  {Akhalwaya}}, \bibinfo {author} {\bibfnamefont {G.}~\bibnamefont
  {Aleksandrowicz}}, \bibinfo {author} {\bibfnamefont {T.}~\bibnamefont
  {Alexander}}, \bibinfo {author} {\bibfnamefont {G.}~\bibnamefont
  {Alexandrowics}}, \bibinfo {author} {\bibfnamefont {E.}~\bibnamefont
  {Arbel}}, \bibinfo {author} {\bibfnamefont {A.}~\bibnamefont {Asfaw}},
  \bibinfo {author} {\bibfnamefont {C.}~\bibnamefont {Azaustre}}, \bibinfo
  {author} {\bibfnamefont {P.}~\bibnamefont {Barkoutsos}}, \bibinfo {author}
  {\bibfnamefont {G.}~\bibnamefont {Barron}}, \bibinfo {author} {\bibfnamefont
  {L.}~\bibnamefont {Bello}}, \bibinfo {author} {\bibfnamefont
  {Y.}~\bibnamefont {Ben-Haim}}, \bibinfo {author} {\bibfnamefont {L.~S.}\
  \bibnamefont {Bishop}}, \bibinfo {author} {\bibfnamefont {S.}~\bibnamefont
  {Bosch}}, \bibinfo {author} {\bibfnamefont {D.}~\bibnamefont {Bucher}},
  \bibinfo {author} {\bibnamefont {CZ}}, \bibinfo {author} {\bibfnamefont
  {F.}~\bibnamefont {Cabrera}}, \bibinfo {author} {\bibfnamefont
  {P.}~\bibnamefont {Calpin}}, \bibinfo {author} {\bibfnamefont
  {L.}~\bibnamefont {Capelluto}}, \bibinfo {author} {\bibfnamefont
  {J.}~\bibnamefont {Carballo}}, \bibinfo {author} {\bibfnamefont {C.-F.}\
  \bibnamefont {Chen}}, \bibinfo {author} {\bibfnamefont {A.}~\bibnamefont
  {Chen}}, \bibinfo {author} {\bibfnamefont {R.}~\bibnamefont {Chen}}, \bibinfo
  {author} {\bibfnamefont {J.~M.}\ \bibnamefont {Chow}}, \bibinfo {author}
  {\bibfnamefont {C.}~\bibnamefont {Claus}}, \bibinfo {author} {\bibfnamefont
  {A.~W.}\ \bibnamefont {Cross}}, \bibinfo {author} {\bibfnamefont {A.~J.}\
  \bibnamefont {Cross}}, \bibinfo {author} {\bibfnamefont {J.}~\bibnamefont
  {Cruz-Benito}}, \bibinfo {author} {\bibfnamefont {C.}~\bibnamefont {Culver}},
  \bibinfo {author} {\bibfnamefont {A.~D.}\ \bibnamefont
  {C{\'o}rcoles-Gonzales}}, \bibinfo {author} {\bibfnamefont {S.}~\bibnamefont
  {Dague}}, \bibinfo {author} {\bibfnamefont {M.}~\bibnamefont {Dartiailh}},
  \bibinfo {author} {\bibfnamefont {A.~R.}\ \bibnamefont {Davila}}, \bibinfo
  {author} {\bibfnamefont {D.}~\bibnamefont {Ding}}, \bibinfo {author}
  {\bibfnamefont {E.}~\bibnamefont {Dumitrescu}}, \bibinfo {author}
  {\bibfnamefont {K.}~\bibnamefont {Dumon}}, \bibinfo {author} {\bibfnamefont
  {I.}~\bibnamefont {Duran}}, \bibinfo {author} {\bibfnamefont
  {P.}~\bibnamefont {Eendebak}}, \bibinfo {author} {\bibfnamefont
  {D.}~\bibnamefont {Egger}}, \bibinfo {author} {\bibfnamefont
  {M.}~\bibnamefont {Everitt}}, \bibinfo {author} {\bibfnamefont {P.~M.}\
  \bibnamefont {Fern{\'a}ndez}}, \bibinfo {author} {\bibfnamefont
  {A.}~\bibnamefont {Frisch}}, \bibinfo {author} {\bibfnamefont
  {A.}~\bibnamefont {Fuhrer}}, \bibinfo {author} {\bibfnamefont
  {J.}~\bibnamefont {Gacon}}, \bibinfo {author} {\bibnamefont {Gadi}}, \bibinfo
  {author} {\bibfnamefont {B.~G.}\ \bibnamefont {Gago}}, \bibinfo {author}
  {\bibfnamefont {J.~M.}\ \bibnamefont {Gambetta}}, \bibinfo {author}
  {\bibfnamefont {L.}~\bibnamefont {Garcia}}, \bibinfo {author} {\bibfnamefont
  {S.}~\bibnamefont {Garion}}, \bibinfo {author} {\bibnamefont {Gawel-Kus}},
  \bibinfo {author} {\bibfnamefont {L.}~\bibnamefont {Gil}}, \bibinfo {author}
  {\bibfnamefont {J.}~\bibnamefont {Gomez-Mosquera}}, \bibinfo {author}
  {\bibfnamefont {S.}~\bibnamefont {de~la Puente~Gonz{\'a}lez}}, \bibinfo
  {author} {\bibfnamefont {D.}~\bibnamefont {Greenberg}}, \bibinfo {author}
  {\bibfnamefont {J.~A.}\ \bibnamefont {Gunnels}}, \bibinfo {author}
  {\bibfnamefont {I.}~\bibnamefont {Haide}}, \bibinfo {author} {\bibfnamefont
  {I.}~\bibnamefont {Hamamura}}, \bibinfo {author} {\bibfnamefont
  {V.}~\bibnamefont {Havlicek}}, \bibinfo {author} {\bibfnamefont
  {J.}~\bibnamefont {Hellmers}}, \bibinfo {author} {\bibfnamefont
  {{\L}.}~\bibnamefont {Herok}}, \bibinfo {author} {\bibfnamefont
  {H.}~\bibnamefont {Horii}}, \bibinfo {author} {\bibfnamefont
  {C.}~\bibnamefont {Howington}}, \bibinfo {author} {\bibfnamefont
  {W.}~\bibnamefont {Hu}}, \bibinfo {author} {\bibfnamefont {S.}~\bibnamefont
  {Hu}}, \bibinfo {author} {\bibfnamefont {H.}~\bibnamefont {Imai}}, \bibinfo
  {author} {\bibfnamefont {T.}~\bibnamefont {Imamichi}}, \bibinfo {author}
  {\bibfnamefont {R.}~\bibnamefont {Iten}}, \bibinfo {author} {\bibfnamefont
  {T.}~\bibnamefont {Itoko}}, \bibinfo {author} {\bibfnamefont
  {A.}~\bibnamefont {Javadi-Abhari}}, \bibinfo {author} {\bibnamefont
  {Jessica}}, \bibinfo {author} {\bibfnamefont {K.}~\bibnamefont {Johns}},
  \bibinfo {author} {\bibfnamefont {N.}~\bibnamefont {Kanazawa}}, \bibinfo
  {author} {\bibfnamefont {A.}~\bibnamefont {Karazeev}}, \bibinfo {author}
  {\bibfnamefont {P.}~\bibnamefont {Kassebaum}}, \bibinfo {author}
  {\bibfnamefont {V.}~\bibnamefont {Krishnan}}, \bibinfo {author}
  {\bibfnamefont {K.}~\bibnamefont {Krsulich}}, \bibinfo {author}
  {\bibfnamefont {G.}~\bibnamefont {Kus}}, \bibinfo {author} {\bibfnamefont
  {R.}~\bibnamefont {LaRose}}, \bibinfo {author} {\bibfnamefont
  {R.}~\bibnamefont {Lambert}}, \bibinfo {author} {\bibfnamefont
  {J.}~\bibnamefont {Latone}}, \bibinfo {author} {\bibfnamefont
  {S.}~\bibnamefont {Lawrence}}, \bibinfo {author} {\bibfnamefont
  {P.}~\bibnamefont {Liu}}, \bibinfo {author} {\bibfnamefont {P.~B.~Z.}\
  \bibnamefont {Mac}}, \bibinfo {author} {\bibfnamefont {Y.}~\bibnamefont
  {Maeng}}, \bibinfo {author} {\bibfnamefont {A.}~\bibnamefont {Malyshev}},
  \bibinfo {author} {\bibfnamefont {J.}~\bibnamefont {Marecek}}, \bibinfo
  {author} {\bibfnamefont {M.}~\bibnamefont {Marques}}, \bibinfo {author}
  {\bibfnamefont {D.}~\bibnamefont {Mathews}}, \bibinfo {author} {\bibfnamefont
  {A.}~\bibnamefont {Matsuo}}, \bibinfo {author} {\bibfnamefont {D.~T.}\
  \bibnamefont {McClure}}, \bibinfo {author} {\bibfnamefont {C.}~\bibnamefont
  {McGarry}}, \bibinfo {author} {\bibfnamefont {D.}~\bibnamefont {McKay}},
  \bibinfo {author} {\bibfnamefont {S.}~\bibnamefont {Meesala}}, \bibinfo
  {author} {\bibfnamefont {A.}~\bibnamefont {Mezzacapo}}, \bibinfo {author}
  {\bibfnamefont {R.}~\bibnamefont {Midha}}, \bibinfo {author} {\bibfnamefont
  {Z.}~\bibnamefont {Minev}}, \bibinfo {author} {\bibfnamefont
  {P.}~\bibnamefont {Murali}}, \bibinfo {author} {\bibfnamefont
  {J.}~\bibnamefont {M{\"u}ggenburg}}, \bibinfo {author} {\bibfnamefont
  {D.}~\bibnamefont {Nadlinger}}, \bibinfo {author} {\bibfnamefont
  {G.}~\bibnamefont {Nannicini}}, \bibinfo {author} {\bibfnamefont
  {P.}~\bibnamefont {Nation}}, \bibinfo {author} {\bibfnamefont
  {Y.}~\bibnamefont {Naveh}}, \bibinfo {author} {\bibnamefont
  {Nick-Singstock}}, \bibinfo {author} {\bibfnamefont {P.}~\bibnamefont
  {Niroula}}, \bibinfo {author} {\bibfnamefont {H.}~\bibnamefont {Norlen}},
  \bibinfo {author} {\bibfnamefont {L.~J.}\ \bibnamefont {O'Riordan}}, \bibinfo
  {author} {\bibfnamefont {S.}~\bibnamefont {Oud}}, \bibinfo {author}
  {\bibfnamefont {D.}~\bibnamefont {Padilha}}, \bibinfo {author} {\bibfnamefont
  {H.}~\bibnamefont {Paik}}, \bibinfo {author} {\bibfnamefont {S.}~\bibnamefont
  {Perriello}}, \bibinfo {author} {\bibfnamefont {A.}~\bibnamefont {Phan}},
  \bibinfo {author} {\bibfnamefont {M.}~\bibnamefont {Pistoia}}, \bibinfo
  {author} {\bibfnamefont {A.}~\bibnamefont {Pozas-iKerstjens}}, \bibinfo
  {author} {\bibfnamefont {V.}~\bibnamefont {Prutyanov}}, \bibinfo {author}
  {\bibfnamefont {J.}~\bibnamefont {P{\'e}rez}}, \bibinfo {author}
  {\bibnamefont {Quintiii}}, \bibinfo {author} {\bibfnamefont {R.}~\bibnamefont
  {Raymond}}, \bibinfo {author} {\bibfnamefont {R.~M.-C.}\ \bibnamefont
  {Redondo}}, \bibinfo {author} {\bibfnamefont {M.}~\bibnamefont {Reuter}},
  \bibinfo {author} {\bibfnamefont {D.~M.}\ \bibnamefont {Rodr{\'\i}guez}},
  \bibinfo {author} {\bibfnamefont {M.}~\bibnamefont {Ryu}}, \bibinfo {author}
  {\bibfnamefont {M.}~\bibnamefont {Sandberg}}, \bibinfo {author}
  {\bibfnamefont {N.}~\bibnamefont {Sathaye}}, \bibinfo {author} {\bibfnamefont
  {B.}~\bibnamefont {Schmitt}}, \bibinfo {author} {\bibfnamefont
  {C.}~\bibnamefont {Schnabel}}, \bibinfo {author} {\bibfnamefont {T.~L.}\
  \bibnamefont {Scholten}}, \bibinfo {author} {\bibfnamefont {E.}~\bibnamefont
  {Schoute}}, \bibinfo {author} {\bibfnamefont {I.~F.}\ \bibnamefont
  {Sertage}}, \bibinfo {author} {\bibfnamefont {Y.}~\bibnamefont {Shi}},
  \bibinfo {author} {\bibfnamefont {A.}~\bibnamefont {Silva}}, \bibinfo
  {author} {\bibfnamefont {Y.}~\bibnamefont {Siraichi}}, \bibinfo {author}
  {\bibfnamefont {S.}~\bibnamefont {Sivarajah}}, \bibinfo {author}
  {\bibfnamefont {J.~A.}\ \bibnamefont {Smolin}}, \bibinfo {author}
  {\bibfnamefont {M.}~\bibnamefont {Soeken}}, \bibinfo {author} {\bibfnamefont
  {D.}~\bibnamefont {Steenken}}, \bibinfo {author} {\bibfnamefont
  {M.}~\bibnamefont {Stypulkoski}}, \bibinfo {author} {\bibfnamefont
  {H.}~\bibnamefont {Takahashi}}, \bibinfo {author} {\bibfnamefont
  {C.}~\bibnamefont {Taylor}}, \bibinfo {author} {\bibfnamefont
  {P.}~\bibnamefont {Taylour}}, \bibinfo {author} {\bibfnamefont
  {S.}~\bibnamefont {Thomas}}, \bibinfo {author} {\bibfnamefont
  {M.}~\bibnamefont {Tillet}}, \bibinfo {author} {\bibfnamefont
  {M.}~\bibnamefont {Tod}}, \bibinfo {author} {\bibfnamefont {E.}~\bibnamefont
  {de~la Torre}}, \bibinfo {author} {\bibfnamefont {K.}~\bibnamefont
  {Trabing}}, \bibinfo {author} {\bibfnamefont {M.}~\bibnamefont {Treinish}},
  \bibinfo {author} {\bibnamefont {TrishaPe}}, \bibinfo {author} {\bibfnamefont
  {W.}~\bibnamefont {Turner}}, \bibinfo {author} {\bibfnamefont
  {Y.}~\bibnamefont {Vaknin}}, \bibinfo {author} {\bibfnamefont {C.~R.}\
  \bibnamefont {Valcarce}}, \bibinfo {author} {\bibfnamefont {F.}~\bibnamefont
  {Varchon}}, \bibinfo {author} {\bibfnamefont {D.}~\bibnamefont {Vogt-Lee}},
  \bibinfo {author} {\bibfnamefont {C.}~\bibnamefont {Vuillot}}, \bibinfo
  {author} {\bibfnamefont {J.}~\bibnamefont {Weaver}}, \bibinfo {author}
  {\bibfnamefont {R.}~\bibnamefont {Wieczorek}}, \bibinfo {author}
  {\bibfnamefont {J.~A.}\ \bibnamefont {Wildstrom}}, \bibinfo {author}
  {\bibfnamefont {R.}~\bibnamefont {Wille}}, \bibinfo {author} {\bibfnamefont
  {E.}~\bibnamefont {Winston}}, \bibinfo {author} {\bibfnamefont {J.~J.}\
  \bibnamefont {Woehr}}, \bibinfo {author} {\bibfnamefont {S.}~\bibnamefont
  {Woerner}}, \bibinfo {author} {\bibfnamefont {R.}~\bibnamefont {Woo}},
  \bibinfo {author} {\bibfnamefont {C.~J.}\ \bibnamefont {Wood}}, \bibinfo
  {author} {\bibfnamefont {R.}~\bibnamefont {Wood}}, \bibinfo {author}
  {\bibfnamefont {S.}~\bibnamefont {Wood}}, \bibinfo {author} {\bibfnamefont
  {J.}~\bibnamefont {Wootton}}, \bibinfo {author} {\bibfnamefont
  {D.}~\bibnamefont {Yeralin}}, \bibinfo {author} {\bibfnamefont
  {J.}~\bibnamefont {Yu}}, \bibinfo {author} {\bibfnamefont {L.}~\bibnamefont
  {Zdanski}}, \bibinfo {author} {\bibnamefont {Zoufalc}}, \bibinfo {author}
  {\bibnamefont {azulehner}}, \bibinfo {author} {\bibnamefont {drholmie}},
  \bibinfo {author} {\bibnamefont {fanizzamarco}}, \bibinfo {author}
  {\bibnamefont {kanejess}}, \bibinfo {author} {\bibnamefont {klinvill}},
  \bibinfo {author} {\bibnamefont {merav aharoni}}, \bibinfo {author}
  {\bibnamefont {ordmoj}}, \bibinfo {author} {\bibnamefont {tigerjack}},
  \bibinfo {author} {\bibnamefont {yang.luh}}, \ and\ \bibinfo {author}
  {\bibnamefont {yotamvakninibm}},\ }\href {\doibase 10.5281/zenodo.2562110}
  {\enquote {\bibinfo {title} {Qiskit: An open-source framework for quantum
  computing},}\ } (\bibinfo {year} {2019})\BibitemShut {NoStop}%
\bibitem [{\citenamefont {Ballance}\ \emph {et~al.}(2016)\citenamefont
  {Ballance}, \citenamefont {Harty}, \citenamefont {Linke}, \citenamefont
  {Sepiol},\ and\ \citenamefont {Lucas}}]{Ballance2016}%
  \BibitemOpen
  \bibfield  {author} {\bibinfo {author} {\bibfnamefont {C.~J.}\ \bibnamefont
  {Ballance}}, \bibinfo {author} {\bibfnamefont {T.~P.}\ \bibnamefont {Harty}},
  \bibinfo {author} {\bibfnamefont {N.~M.}\ \bibnamefont {Linke}}, \bibinfo
  {author} {\bibfnamefont {M.~A.}\ \bibnamefont {Sepiol}}, \ and\ \bibinfo
  {author} {\bibfnamefont {D.~M.}\ \bibnamefont {Lucas}},\ }\href {\doibase
  10.1103/PhysRevLett.117.060504} {\bibfield  {journal} {\bibinfo  {journal}
  {Phys. Rev. Lett.}\ }\textbf {\bibinfo {volume} {117}},\ \bibinfo {pages}
  {060504} (\bibinfo {year} {2016})}\BibitemShut {NoStop}%
\bibitem [{\citenamefont {Monz}\ \emph {et~al.}(2011)\citenamefont {Monz},
  \citenamefont {Schindler}, \citenamefont {Barreiro}, \citenamefont {Chwalla},
  \citenamefont {Nigg}, \citenamefont {Coish}, \citenamefont {Harlander},
  \citenamefont {H\"ansel}, \citenamefont {Hennrich},\ and\ \citenamefont
  {Blatt}}]{Monz2011}%
  \BibitemOpen
  \bibfield  {author} {\bibinfo {author} {\bibfnamefont {T.}~\bibnamefont
  {Monz}}, \bibinfo {author} {\bibfnamefont {P.}~\bibnamefont {Schindler}},
  \bibinfo {author} {\bibfnamefont {J.~T.}\ \bibnamefont {Barreiro}}, \bibinfo
  {author} {\bibfnamefont {M.}~\bibnamefont {Chwalla}}, \bibinfo {author}
  {\bibfnamefont {D.}~\bibnamefont {Nigg}}, \bibinfo {author} {\bibfnamefont
  {W.~A.}\ \bibnamefont {Coish}}, \bibinfo {author} {\bibfnamefont
  {M.}~\bibnamefont {Harlander}}, \bibinfo {author} {\bibfnamefont
  {W.}~\bibnamefont {H\"ansel}}, \bibinfo {author} {\bibfnamefont
  {M.}~\bibnamefont {Hennrich}}, \ and\ \bibinfo {author} {\bibfnamefont
  {R.}~\bibnamefont {Blatt}},\ }\href {\doibase 10.1103/PhysRevLett.106.130506}
  {\bibfield  {journal} {\bibinfo  {journal} {Phys. Rev. Lett.}\ }\textbf
  {\bibinfo {volume} {106}},\ \bibinfo {pages} {130506} (\bibinfo {year}
  {2011})}\BibitemShut {NoStop}%
\bibitem [{\citenamefont {Cross}\ \emph {et~al.}(2017)\citenamefont {Cross},
  \citenamefont {Bishop}, \citenamefont {Smolin},\ and\ \citenamefont
  {Gambetta}}]{openqasm}%
  \BibitemOpen
  \bibfield  {author} {\bibinfo {author} {\bibfnamefont {A.~W.}\ \bibnamefont
  {Cross}}, \bibinfo {author} {\bibfnamefont {L.}~\bibnamefont {Bishop}},
  \bibinfo {author} {\bibfnamefont {J.}~\bibnamefont {Smolin}}, \ and\ \bibinfo
  {author} {\bibfnamefont {J.~M.}\ \bibnamefont {Gambetta}},\ }\href
  {https://arxiv.org/abs/1707.03429} {\bibfield  {journal} {\bibinfo  {journal}
  {arXiv preprint}\ }\textbf {\bibinfo {volume} {arXiv:1707.03429}} (\bibinfo
  {year} {2017})}\BibitemShut {NoStop}%
\bibitem [{\citenamefont {Cuthill}\ and\ \citenamefont
  {McKee}(1969)}]{Cuthill69}%
  \BibitemOpen
  \bibfield  {author} {\bibinfo {author} {\bibfnamefont {E.}~\bibnamefont
  {Cuthill}}\ and\ \bibinfo {author} {\bibfnamefont {J.}~\bibnamefont
  {McKee}},\ }\href {\doibase 10.1145/800195.805928} {\bibfield  {journal}
  {\bibinfo  {journal} {Proc. 24th Nat. Conf. ACM}\ ,\ \bibinfo {pages} {157}}
  (\bibinfo {year} {1969})}\BibitemShut {NoStop}%
\bibitem [{\citenamefont {George}\ and\ \citenamefont {Liu}(1981)}]{George81}%
  \BibitemOpen
  \bibfield  {author} {\bibinfo {author} {\bibfnamefont {A.}~\bibnamefont
  {George}}\ and\ \bibinfo {author} {\bibfnamefont {J.~W.}\ \bibnamefont
  {Liu}},\ }\href@noop {} {\emph {\bibinfo {title} {{C}omputer {S}olution of
  {L}arge {S}parse {P}ositive {D}efinite {S}ystems}}}\ (\bibinfo  {publisher}
  {Prentice-Hall},\ \bibinfo {year} {1981})\BibitemShut {NoStop}%
\bibitem [{\citenamefont {Chan}\ and\ \citenamefont {George}(1980)}]{Chan80}%
  \BibitemOpen
  \bibfield  {author} {\bibinfo {author} {\bibfnamefont {W.~M.}\ \bibnamefont
  {Chan}}\ and\ \bibinfo {author} {\bibfnamefont {A.}~\bibnamefont {George}},\
  }\href {\doibase 10.1007/BF01933580} {\bibfield  {journal} {\bibinfo
  {journal} {BIT}\ }\textbf {\bibinfo {volume} {20}},\ \bibinfo {pages} {8}
  (\bibinfo {year} {1980})}\BibitemShut {NoStop}%
\bibitem [{\citenamefont {Kraus}\ and\ \citenamefont
  {Cirac}(2001)}]{kraus_optimal_2001}%
  \BibitemOpen
  \bibfield  {author} {\bibinfo {author} {\bibfnamefont {B.}~\bibnamefont
  {Kraus}}\ and\ \bibinfo {author} {\bibfnamefont {J.~I.}\ \bibnamefont
  {Cirac}},\ }\href {\doibase 10.1103/PhysRevA.63.062309} {\bibfield  {journal}
  {\bibinfo  {journal} {Physical Review A}\ }\textbf {\bibinfo {volume} {63}},\
  \bibinfo {pages} {062309} (\bibinfo {year} {2001})}\BibitemShut {NoStop}%
\bibitem [{\citenamefont {Khaneja}\ \emph {et~al.}(2001)\citenamefont
  {Khaneja}, \citenamefont {Brockett},\ and\ \citenamefont
  {Glaser}}]{khaneja_time_2001}%
  \BibitemOpen
  \bibfield  {author} {\bibinfo {author} {\bibfnamefont {N.}~\bibnamefont
  {Khaneja}}, \bibinfo {author} {\bibfnamefont {R.}~\bibnamefont {Brockett}}, \
  and\ \bibinfo {author} {\bibfnamefont {S.~J.}\ \bibnamefont {Glaser}},\
  }\href {\doibase 10.1103/PhysRevA.63.032308} {\bibfield  {journal} {\bibinfo
  {journal} {Physical Review A}\ }\textbf {\bibinfo {volume} {63}},\ \bibinfo
  {pages} {032308} (\bibinfo {year} {2001})}\BibitemShut {NoStop}%
\bibitem [{\citenamefont {Watts}\ \emph {et~al.}(2015)\citenamefont {Watts},
  \citenamefont {Vala}, \citenamefont {M{\"u}ller}, \citenamefont {Calarco},
  \citenamefont {Whaley}, \citenamefont {Reich}, \citenamefont {Goerz},\ and\
  \citenamefont {Koch}}]{watts_optimizing_2015}%
  \BibitemOpen
  \bibfield  {author} {\bibinfo {author} {\bibfnamefont {P.}~\bibnamefont
  {Watts}}, \bibinfo {author} {\bibfnamefont {J.}~\bibnamefont {Vala}},
  \bibinfo {author} {\bibfnamefont {M.~M.}\ \bibnamefont {M{\"u}ller}},
  \bibinfo {author} {\bibfnamefont {T.}~\bibnamefont {Calarco}}, \bibinfo
  {author} {\bibfnamefont {K.~B.}\ \bibnamefont {Whaley}}, \bibinfo {author}
  {\bibfnamefont {D.~M.}\ \bibnamefont {Reich}}, \bibinfo {author}
  {\bibfnamefont {M.~H.}\ \bibnamefont {Goerz}}, \ and\ \bibinfo {author}
  {\bibfnamefont {C.~P.}\ \bibnamefont {Koch}},\ }\href {\doibase
  10.1103/PhysRevA.91.062306} {\bibfield  {journal} {\bibinfo  {journal}
  {Physical Review A}\ }\textbf {\bibinfo {volume} {91}},\ \bibinfo {pages}
  {062306} (\bibinfo {year} {2015})}\BibitemShut {NoStop}%
\bibitem [{\citenamefont {Vatan}\ and\ \citenamefont
  {Williams}(2004)}]{vatan_optimal_2004}%
  \BibitemOpen
  \bibfield  {author} {\bibinfo {author} {\bibfnamefont {F.}~\bibnamefont
  {Vatan}}\ and\ \bibinfo {author} {\bibfnamefont {C.}~\bibnamefont
  {Williams}},\ }\href {\doibase 10.1103/PhysRevA.69.032315} {\bibfield
  {journal} {\bibinfo  {journal} {Physical Review A}\ }\textbf {\bibinfo
  {volume} {69}},\ \bibinfo {pages} {032315} (\bibinfo {year}
  {2004})}\BibitemShut {NoStop}%
\bibitem [{\citenamefont {Vidal}\ and\ \citenamefont
  {Dawson}(2004)}]{vidal_universal_2004}%
  \BibitemOpen
  \bibfield  {author} {\bibinfo {author} {\bibfnamefont {G.}~\bibnamefont
  {Vidal}}\ and\ \bibinfo {author} {\bibfnamefont {C.~M.}\ \bibnamefont
  {Dawson}},\ }\href {\doibase 10.1103/PhysRevA.69.010301} {\bibfield
  {journal} {\bibinfo  {journal} {Physical Review A}\ }\textbf {\bibinfo
  {volume} {69}},\ \bibinfo {pages} {010301} (\bibinfo {year}
  {2004})}\BibitemShut {NoStop}%
\bibitem [{\citenamefont {Zhang}\ \emph {et~al.}(2005)\citenamefont {Zhang},
  \citenamefont {Ye},\ and\ \citenamefont {Guo}}]{zhang_conditions_2005}%
  \BibitemOpen
  \bibfield  {author} {\bibinfo {author} {\bibfnamefont {Y.-S.}\ \bibnamefont
  {Zhang}}, \bibinfo {author} {\bibfnamefont {M.-Y.}\ \bibnamefont {Ye}}, \
  and\ \bibinfo {author} {\bibfnamefont {G.-C.}\ \bibnamefont {Guo}},\ }\href
  {\doibase 10.1103/PhysRevA.71.062331} {\bibfield  {journal} {\bibinfo
  {journal} {Physical Review A}\ }\textbf {\bibinfo {volume} {71}},\ \bibinfo
  {pages} {062331} (\bibinfo {year} {2005})}\BibitemShut {NoStop}%
\bibitem [{\citenamefont {Watts}\ \emph {et~al.}(2013)\citenamefont {Watts},
  \citenamefont {O'Connor}, \citenamefont {Vala}, \citenamefont {Watts},
  \citenamefont {O'Connor},\ and\ \citenamefont {Vala}}]{watts_metric_2013}%
  \BibitemOpen
  \bibfield  {author} {\bibinfo {author} {\bibfnamefont {P.}~\bibnamefont
  {Watts}}, \bibinfo {author} {\bibfnamefont {M.}~\bibnamefont {O'Connor}},
  \bibinfo {author} {\bibfnamefont {J.}~\bibnamefont {Vala}}, \bibinfo {author}
  {\bibfnamefont {P.}~\bibnamefont {Watts}}, \bibinfo {author} {\bibfnamefont
  {M.}~\bibnamefont {O'Connor}}, \ and\ \bibinfo {author} {\bibfnamefont
  {J.}~\bibnamefont {Vala}},\ }\href {\doibase 10/f44m4x} {\bibfield  {journal}
  {\bibinfo  {journal} {Entropy}\ }\textbf {\bibinfo {volume} {15}},\ \bibinfo
  {pages} {1963} (\bibinfo {year} {2013})}\BibitemShut {NoStop}%
\bibitem [{\citenamefont {Musz}\ \emph {et~al.}(2013)\citenamefont {Musz},
  \citenamefont {Ku{\'s}},\ and\ \citenamefont
  {{\.Z}yczkowski}}]{musz_unitary_2013}%
  \BibitemOpen
  \bibfield  {author} {\bibinfo {author} {\bibfnamefont {M.}~\bibnamefont
  {Musz}}, \bibinfo {author} {\bibfnamefont {M.}~\bibnamefont {Ku{\'s}}}, \
  and\ \bibinfo {author} {\bibfnamefont {K.}~\bibnamefont {{\.Z}yczkowski}},\
  }\href {\doibase 10/gfgnq7} {\bibfield  {journal} {\bibinfo  {journal}
  {Physical Review A}\ }\textbf {\bibinfo {volume} {87}},\ \bibinfo {pages}
  {022111} (\bibinfo {year} {2013})}\BibitemShut {NoStop}%
\end{thebibliography}
\end{document}